\documentclass[twocolumn]{aastex62}%,trackchanges
\usepackage{amsmath,color,textcomp,url,graphicx,subfigure}
\usepackage{listings}

\newcommand{\initialobjects}{3158}
\newcommand{\knownobjects}{1399}%dont forget to remove NKastner
\newcommand{\kastnerobjects}{12}
\newcommand{\knownpluskastnerobjects}{1411}
\newcommand{\comovingsystems}{104}%28 lost from CMDs
\newcommand{\rejectedbyrv}{18}
\newcommand{\literaturervs}{36}
\newcommand{\rejectedbyliteraturerv}{6}
\newcommand{\rejectedbycomovingrv}{12}
\newcommand{\strengthenedbycomovingrv}{28}%1 lost from CMDs
\newcommand{\comovingrvs}{40}%1 lost from CMDs
\newcommand{\rejectedbyvisual}{504}%dont forget to count the WDs
\newcommand{\rejectedbycmd}{327}
\newcommand{\finalcandidates}{898}%previously 1225, before the CMD rejections
\newcommand{\browndwarfs}{111}%36 CMD-rejected
\newcommand{\ldwarfs}{19}%none CMD-rejected

\newcommand{\kms}{\hbox{km\,s$^{-1}$}}

\newcommand{\masyr}{$\mathrm{mas}\,\mathrm{yr}^{-1}$}

\received{2018 May 7}
\revised{2018 May 26}
\submitjournal{ApJ}

\shorttitle{Young Associations in \emph{Gaia} DR2}
\shortauthors{Gagn\'e et al.}

\begin{document}

\title{BANYAN. XIII. A FIRST LOOK AT NEARBY YOUNG ASSOCIATIONS WITH \emph{GAIA} DATA RELEASE 2}

\author[0000-0002-2592-9612]{Jonathan Gagn\'e}
\affiliation{Carnegie Institution of Washington DTM, 5241 Broad Branch Road NW, Washington, DC~20015, USA}
\affiliation{NASA Sagan Fellow}
\email{jgagne@carnegiescience.edu}
\author[0000-0001-6251-0573]{Jacqueline K. Faherty}
\affiliation{Department of Astrophysics, American Museum of Natural History, Central Park West at 79th St., New York, NY 10024, USA}

\begin{abstract}

In this paper we examine the nearest 100\,pc entries in the data release 2 of \emph{Gaia} to identify previously unrecognized candidate members in young associations. We analyze 695\,952 stars with the BANYAN~$\Sigma$ Bayesian classification software and discover \finalcandidates\  new high-likelihood candidate members that span a wide range in properties, from spectral types B9 to L2, including \comovingsystems\ co-moving systems, \browndwarfs\ brown dwarfs and 31 new bona fide members. Our sample is mostly composed of highly active M dwarfs and will be crucial to examine the low-mass end of the initial mass function of young associations. Our sample includes new candidate members near the Galactic plane where previous surveys suffered from a high rate of contamination. This paper represents the first step towards a full reassessment of young associations in the Solar neighborhood with the second data release of the \emph{Gaia} mission.

\end{abstract}


\section{INTRODUCTION}\label{sec:intro}

The recent Data Release 2 of the \emph{Gaia} mission (\emph{Gaia}~DR2 hereafter; \added{\citealt{GaiaCollaboration:2018io,Lindegren:2018gy}})\footnote{\added{See also \cite{Luri:2018eu}, \cite{Mignard:2018bj}, \cite{Babusiaux:2018di},\cite{Sartoretti:2018jm}, \cite{Soubiran:2018fz}, \cite{Cropper:2018jx}, \cite{Evans:2018cj}, \cite{Hambly:2018gr}, and \cite{Riello:2018bo} for relevant calibration.}} on 2018 April 25 presented a catalog of $\sim$\,1.3 billion trigonometric distances to stars in the Milky Way, a more than 600-fold improvement over the number of sources in the Tycho--\emph{Gaia} Astrometric Solution \citep{2016AA...595A...2G} that came out less than a year ago, and a more than 10\,000-fold improvement over that of the Hipparcos mission \citep{1997AA...323L..49P}, which shaped our understanding of the Solar neighborhood for the past few decades. \emph{Gaia}~DR2 revolutionizes the quantity and quality of stellar kinematics data that are immediately available, and it will have a profound impact on our understanding of Galactic kinematics, among many other things.

In this paper, we examine the 27 nearest known young associations within 150\,pc of the Sun (e.g., the AB~Doradus and $\beta$~Pictoris moving groups, the TW~Hya and Hyades associations; \citealp{2004ARAA..42..685Z,2008hsf2.book..757T}). These associations and their global properties are listed in Table~\ref{tab:nyas}, and are described in more detail in \cite{2018ApJ...856...23G}. A detailed analysis of their properties and known members based on \emph{Gaia}~DR2 will be the subject of separate papers. The ages of these associations span a few Myr to almost a Gyr, and they thus provide a window into the star-formation history of the Solar neighborhood as well as important astrophysical laboratories to understand how the properties of stars, substellar objects and exoplanets evolve with time. Previous searches for new low-mass members in these young associations (e.g., \citealt{2011ApJ...727...62R,2012ApJ...758...56S,2013ApJ...762...88M,2014AJ....147..146K,2017AJ....154...69S}) identified only a fraction of the mid-to-late M dwarfs because of their faintness and the need to obtain follow-up trigonometric parallaxes and radial velocities for a large number of objects (the initial mass function peaks around $\sim$\,M3). \emph{Gaia}~DR2 now opens the door to a search down to the substellar domain (spectral type $\sim$\,L2) with an \replaced{unprecendeted}{unprecedented} efficiency, as the immediate availability of trigonometric parallaxes makes it possible to cut down the number of contaminants (mostly unrelated background field stars) by an order of magnitude (e.g., see \citealt{2018ApJ...856...23G}).

\startlongtable
\tablewidth{0.985\textwidth}
\begin{deluxetable*}{lccccccc}
\tablecolumns{7}
\tablecaption{Nearby young associations considered here.\label{tab:nyas}}
\tablehead{\colhead{Group} & \colhead{$\left<\varpi\right>$\tablenotemark{a}} & \colhead{$\left<\nu\right>$\tablenotemark{b}} & \colhead{$S_{\rm spa}$\tablenotemark{c}} & \colhead{$S_{\rm kin}$\tablenotemark{d}} & \colhead{Age} & \colhead{Ref.}\\
\colhead{Name} & \colhead{(pc)} & \colhead{(\kms)} & \colhead{(pc)} & \colhead{(\kms)} & \colhead{(Myr)} & \colhead{} }
\startdata
118TAU & $100 \pm 10$ & $14 \pm 2$ & 3.4 & 2.1 & $\sim$\,10 & 1\\
ABDMG & $30_{-10}^{+20}$ & $10_{-20}^{+10}$ & 19.0 & 1.4 & $149_{-19}^{+51}$ & 2\\
$\beta$PMG & $30_{-10}^{+20}$ & $10 \pm 10$ & 14.8 & 1.4 & $24 \pm 3$ & 2\\
CAR & $60 \pm 20$ & $20 \pm 2$ & 11.8 & 0.8 & $45_{-7}^{+11}$ & 2\\
CARN & $30 \pm 20$ & $15_{-10}^{+7}$ & 14.0 & 2.1 & $\sim$\,200 & 3\\
CBER & $85_{-5}^{+4}$ & $-0.1 \pm 0.8$ & 3.6 & 0.5 & $562_{-84}^{+98}$ & 4\\
COL & $50 \pm 20$ & $21_{-8}^{+3}$ & 15.8 & 0.9 & $42_{-4}^{+6}$ & 2\\
EPSC & $102 \pm 4$ & $14 \pm 3$ & 2.8 & 1.8 & $3.7_{-1.4}^{+4.6}$ & 5\\
HYA & $42 \pm 7$ & $39_{-4}^{+3}$ & 4.5 & 1.2 & $750 \pm 100$ & 6\\
LCC & $110 \pm 10$ & $14 \pm 5$ & 11.6 & 2.2 & $15 \pm 3$ & 7\\
OCT & $130_{-20}^{+30}$ & $8_{-9}^{+8}$ & 22.4 & 1.3 & $35 \pm 5$ & 8\\
PL8 & $130 \pm 10$ & $22 \pm 2$ & 5.0 & 1.1 & $\sim$\,60 & 9\\
TAU & $120 \pm 10$ & $16 \pm 3$ & 10.7 & 3.6 & 1--2 & 10\\
THA & $46_{-6}^{+8}$ & $9_{-6}^{+5}$ & 9.1 & 0.8 & $45 \pm 4$ & 2\\
THOR & $96 \pm 2$ & $19 \pm 3$ & 3.9 & 2.1 & $22_{-3}^{+4}$ & 2\\
TWA & $60 \pm 10$ & $10 \pm 3$ & 6.6 & 1.5 & $10 \pm 3$ & 2\\
UCL & $130 \pm 20$ & $5 \pm 5$ & 17.4 & 2.5 & $16 \pm 2$ & 7\\
UMA & $25.4_{-0.7}^{+0.8}$ & $-12 \pm 3$ & 1.2 & 1.3 & $414 \pm 23$ & 11\\
USCO & $130 \pm 20$ & $-5 \pm 4$ & 9.9 & 2.8 & $10 \pm 3$ & 7\\
XFOR & $100 \pm 6$ & $19 \pm 2$ & 2.6 & 1.3 & $\sim$\,500 & 12\\
\enddata
\tablenotetext{a}{Peak of distance distribution and $\pm$1$\sigma$ range.}
\tablenotetext{b}{Peak of radial velocity distribution and $\pm$1$\sigma$ range.}
\tablenotetext{c}{Characteristic spatial scale in $XYZ$ space.}
\tablenotetext{d}{Characteristic kinematic scale in $UVW$ space.}
%\tablewidth{0.4\textwidth}
\tablecomments{The full names of young associations are: 118~Tau (118TAU), AB~Doradus (ABDMG), $\beta$~Pictoris ($\beta$PMG), Carina (CAR), Carina-Near (CARN), Coma Berenices (CBER), Columba (COL), $\epsilon$~Chamaeleontis (EPSC), the Hyades cluster (HYA), Lower Centaurus Crux (LCC), Octans (OCT), Platais~8 (PL8), the Tucana-Horologium association (THA), 32~Orionis (THOR), TW~Hya (TWA), Upper Centaurus Lupus (UCL), the core of the Ursa~Major cluster (UMA), Upper~Scorpius (USCO), Taurus (TAU) and $\chi^1$~For (XFOR).}
\tablerefs{(1)~\citealt{mamajek118tau}; (2)~\citealt{2015MNRAS.454..593B}; (3)~\citealt{2006ApJ...649L.115Z}; (4)~\citealt{2014AA...566A.132S}; (5)~\citealt{2013MNRAS.435.1325M}; (6)~\citealt{2015ApJ...807...24B}; (7)~\citealt{2016MNRAS.461..794P}; (8)~\citealt{2015MNRAS.447.1267M}; (9)~\citealt{1998AJ....116.2423P}; (10)~\citealt{1995ApJS..101..117K}; (11)~\citealt{2015AAS...22511203J}; (12)~\citealt{2010AA...514A..81P}.}
\end{deluxetable*}

\clearpage

In this work, we use the 100\,pc sample of \emph{Gaia}~DR2 to recover \finalcandidates\ new candidate members (mostly mid-M dwarfs) that were never identified as such in the literature. In Section~\ref{sec:sample}, we describe the sample selection based on the nearest 100\,pc entries of \emph{Gaia}~DR2 and the BANYAN~$\Sigma$ Bayesian classifier, and we describe our further validation of the targets as well as our literature search in Section~\ref{sec:lit}. The general properties of the resulting set of new candidate members are described in Section~\ref{sec:discussion}, along with those of the co-moving systems that we identified. We conclude in Section~\ref{sec:conclusion}.

\section{SAMPLE SELECTION}\label{sec:sample}

We selected all \emph{Gaia}~DR2 sources within 100\,pc of the Sun with a parallax measurement at least three times as large as its measurement error so as to identify new candidate members in young associations based on robust parallax measurements. \added{We used a generous criterion on the parallax quality to avoid rejecting faint candidate members that may correspond to young brown dwarfs}. This sample was downloaded from the \emph{Gaia}~DR2 archive\footnote{\url{https://archives.esac.esa.int/gaia}} with the following SQL query:
\begin{lstlisting}[language=SQL,basicstyle=\ttfamily\small]
PARALLAX>10 AND PARALLAX/PARALLAX_ERROR>=3,
\end{lstlisting}
\noindent which returned 695\,952 entries. All entries were analyzed with the BANYAN~$\Sigma$ Bayesian classification algorithm \citep{2018ApJ...856...23G} to identify those for which the proper motion and trigonometric parallax are consistent with membership in a known young association. BANYAN~$\Sigma$ uses spatial and kinematic models of the 27 young associations within 150\,pc as well as field stars within 300\,pc to derive membership probabilities in each young association based on the sky position, proper motion, radial velocity and trigonometric distances of the targets using Bayesian inference. Radial velocities and trigonometric distances are optional, and BANYAN~$\Sigma$ can calculate probabilities in their absence by taking a marginalization integral of the membership probability density over all possible values. BANYAN~$\Sigma$ is computationally efficient and achieves a more accurate classification performance than previous tools in the literature in part because it is based on an anlaytical solution of the marginalization integrals (see \citealt{2018ApJ...856...23G} for more detail).

\begin{figure}
	\centering
	\includegraphics[width=0.465\textwidth]{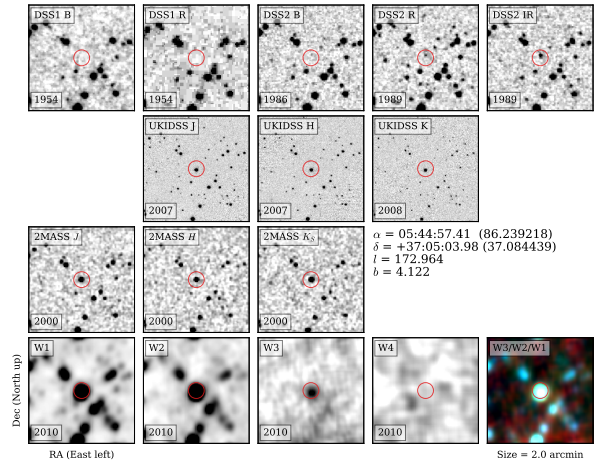}
	\caption{Finder chart for 2MASS~J05445741+3705039, a new Galactic plane $\sim$\,L1 candidate member of the ABDMG. The red circle displays its expected position at epoch 2000 based on the \emph{Gaia} astrometric solution. Its large proper motion ($\mu_\alpha\cos\delta = 6.6 \pm 0.9$\,\masyr, $\mu_\delta = -230.8 \pm 0.8$\,\masyr) can clearly be seen from the 1989 DSS2 infrared image and the $\sim$\,2007 UKIDSS images, and its non-detection at bluer wavelengths is consistent with its estimated spectral type. See Section~\ref{sec:lit} for more detail.}
	\label{fig:fchart1}
\end{figure}
\begin{figure}
	\centering
	\includegraphics[width=0.465\textwidth]{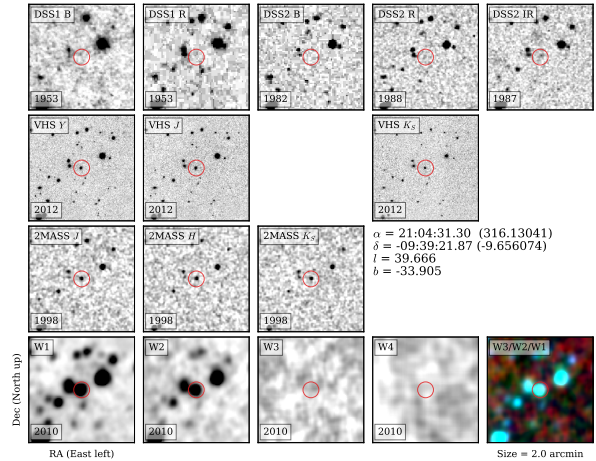}
	\caption{Finder chart for 2MASS~J21043128--0939217 , a new $\sim$\,L3 candidate member of the $\beta$PMG. The red circle displays its expected position at epoch 2000 based on the \emph{Gaia} astrometric solution. Its proper motion can be seen from the marginal detection in the 1989 DSS2 infrared image as well as the slightly off-center sources in the VHS and \emph{WISE} images, and is consistent with the \emph{Gaia} proper motion ($\mu_\alpha\cos\delta = 59 \pm 3$\,\masyr,$\mu_\delta = -58 \pm 2$\,\masyr). Its non-detection at optical wavelengths is also consistent with its estimated spectral type. See Section~\ref{sec:lit} for more detail.}
	\label{fig:fchart2}
\end{figure}

We found all 3,510 entries with a total young association Bayesian membership probabily above 90\% that are separated by less than 5\,\kms\ from their best-matching young association in $UVW$ space (727 of which have a \emph{Gaia}~DR2 radial velocity measurement). In the absence of a radial velocity measurement, BANYAN~$\Sigma$ provides the optimal valye that would minimize the distance between a target and its most probable young association in spatial-kinematic $XYZUVW$ space. The 90\% probability threshold is associated with a recovery rate of only 50\% of the bona fide members \citep{2018ApJ...856...23G}, and was selected as a first pass to identify only the most unambiguous members of young associations in \emph{Gaia}~DR2 while minimizing the rate of contamination from random field interlopers.

\section{LITERATURE SEARCH AND TARGET VALIDATION}\label{sec:lit}

\cite{2018ApJ...856...23G} compiled a list of candidate or bona fide members of young associations. Before our sample was analyzed in more detail, we cross-matched it with this list and removed from our sample the \knownobjects\ that were found in both lists. The known young brown dwarfs will be discussed in J.~K.~Faherty et al. (in preparation), and the known stellar members will be discussed in J.~Gagn\'e et al. (in preparation) along with an update of the spatial-kinematic models of the young associations and the field used in BANYAN~$\Sigma$. We also ignored \kastnerobjects\ K-type stars that will be discussed by M. Chalifour, A. Binks, J. Kastner et al. (in preparation).

\begin{figure}
	\centering
	\includegraphics[width=0.465\textwidth]{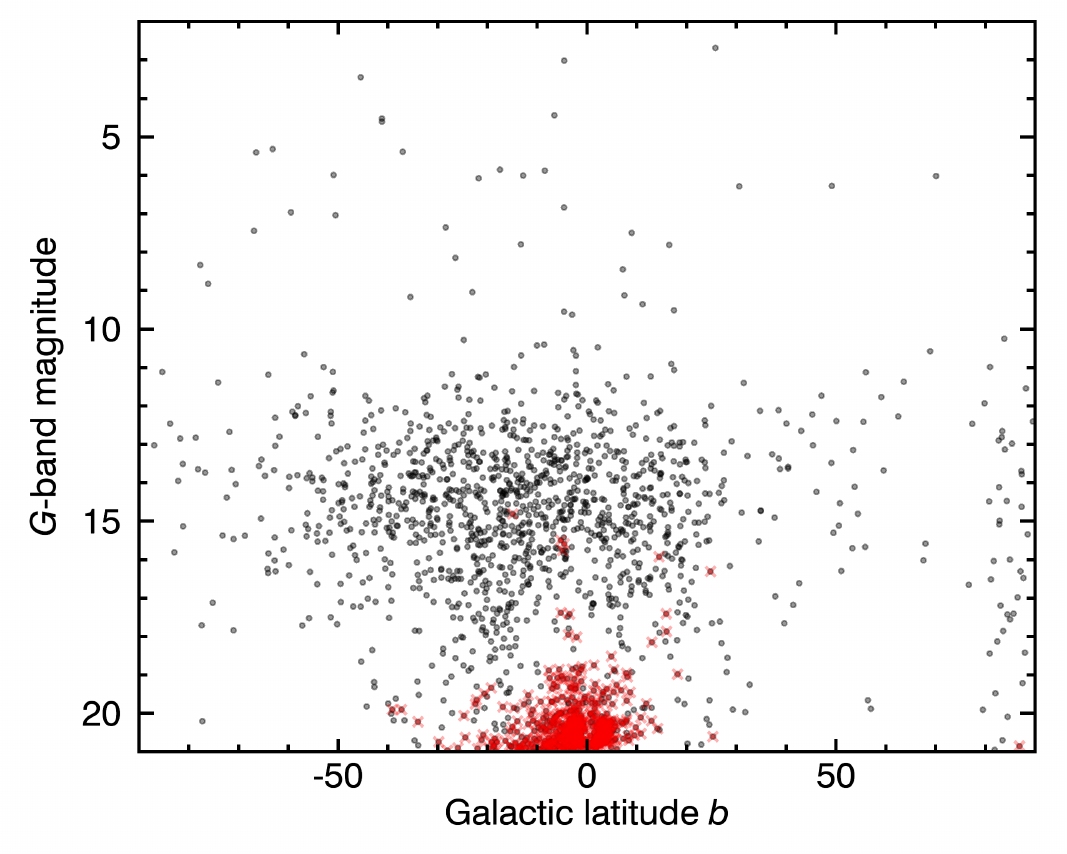}
	\caption{Relative \emph{Gaia} $G$--band magnitude versus Galactic latitude for the full sample (black circles) and unphysical objects rejected by the finder charts examination (red crosses). Some \emph{Gaia}~DR2 parallax solutions correspond to unphysical objects, and are mainly caused by cross-match confusion of faint objects near the Galactic plane. See Section~\ref{sec:lit} for more detail.}
	\label{fig:badsols}
\end{figure}

\begin{figure}
	\centering
	\includegraphics[width=0.465\textwidth]{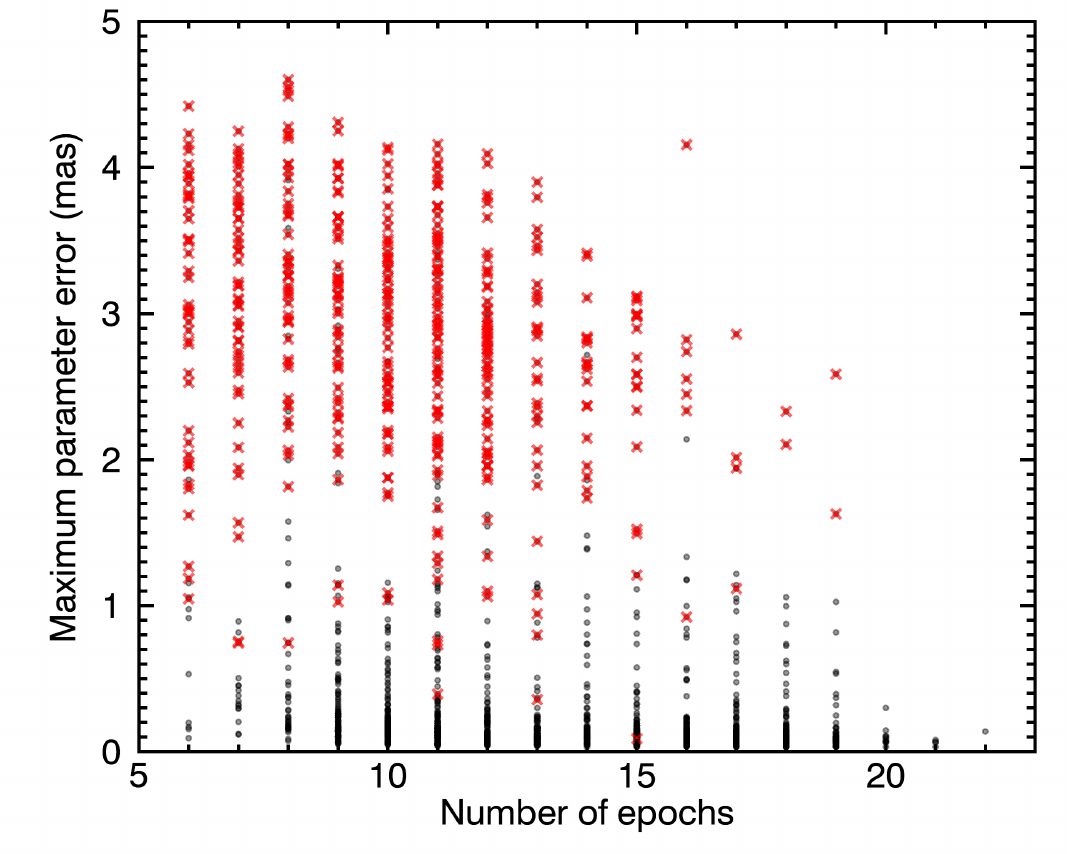}
	\caption{Maximum measurement error on all parameters of the \emph{Gaia}~DR2 astrometric solution (\texttt{ASTROMETRIC\_SIGMA5D\_MAX}) versus the number of visibility windows used in the solution (\texttt{VISIBILITY\_PERIODS\_USED}), for the full sample (black circles) and those rejected as unphysical entries by a visual examination of the finder charts (red crosses). As expected, entries with less epochs and larger error bars are more prone to correspond to unphysical objects, but a simple rejection criterion based on these parameters would also reject some good solutions. See Section~\ref{sec:lit} for more detail.}
	\label{fig:badsols2}
\end{figure}

Several parallax solutions near the Galactic plane in \emph{Gaia}~DR2 suffer from cross-matching confusion, especially for faint targets (\citealt{Lindegren:2018gy}; see also \citealt{2018arXiv180501573F}). In order to assess which \emph{Gaia} entries corresponded to physical objects, we built finder charts with all available data from the Digitized Sky Survey, SDSS \citep{2015ApJS..219...12A}, 2MASS \citep{2006AJ....131.1163S}, \emph{WISE} \citep{2010AJ....140.1868W}, Pan-STARRS \citep{2016arXiv161205560C}, VHS \citep{2013Msngr.154...35M} and UKIDSS \citep{2007MNRAS.379.1599L} with the \texttt{finder\_charts.py} Python package \citep{zenodofindercharts}\footnote{Available at \url{https://github.com/jgagneastro/finder_charts} and at \url{http://www.astro.umontreal.ca/~gagne/finderchart.php} as a web tool.}. We visually examined the 2195 charts to confirm the non-zero proper motion of the targets and to verify that their colors were consistent with their absolute \emph{Gaia} $G$-band magnitudes (e.g., see Figures~\ref{fig:fchart1} and \ref{fig:fchart2}). This step allowed us to remove \rejectedbyvisual\ contaminants that corresponded to unphysical entries in the \emph{Gaia} catalog. Their distribution in relative $G$--band magnitude versus Galactic latitude is displayed in Figure~\ref{fig:badsols} compared to the full input sample, which demonstrates that most of the unphysical \emph{Gaia}~DR2 parallax solutions are caused by cross-match confusion of faint objects near the Galactic plane. Most of them can also be rejected from an inspection of the \texttt{ASTROMETRIC\_SIGMA5D\_MAX} and \texttt{VISIBILITY\_PERIODS\_USED} flags in the \emph{Gaia}~DR2 catalog, which respectively correspond to the maximal measurement error of all astrometric solution parameters (in mas), and to the number of epochs that were used in the solution (see Figure~\ref{fig:badsols2}). However, a simple rejection filter based on these criteria would also inevitably reject some good solutions.

\added{A total of 250/261 initial entries with \emph{Gaia}~DR2 parallax measurements in the range 3--8 times larger than the measurement error were rejected by our visual inspection of the finder charts. A large fraction of low-precision parallaxes thus corresponded to unphysical \emph{Gaia}~DR2 entries, but we keep the remaining 11 low-precision entries because they likely correspond to valid late-M or early-L candidate members.}

Our literature search revealed that \literaturervs\ of our new candidate members without a \emph{Gaia}~DR2 radial velocity have such measurements in the literature, \rejectedbyliteraturerv\ of which rejected the membership hypothesis.

There are four high-probability candidate members of $\beta$PMG and ABDMG in our sample that appear in the literature as candidates of the more distant PLE and HYA associations and the TAU and Sco-Cen star-forming regions. These objects were rejected from our sample as a consequence of this, but we provide below a short discussion of their membership.

Both 2MASS J17513421--4854558 ($\approx$\,M1) and 2MASS J17194204--4615275 (M2; \citealp{2016AA...594A..63G}) were identified as young candidate members of USCO with Li absorption (290\,m\AA\ and 520\,m\AA\ respectively) by \cite{2012AJ....144....8S}. However, their \emph{Gaia}~DR2 trigonometric distances ($66.5 \pm 1.2$\,pc and $53.0 \pm 0.2$\,pc respectively) preclude membership in USCO, and strongly favors membership in the $\beta$PMG. In both cases, radial velocity measurements are still needed to confirm their membership, but the \emph{Gaia}~DR2 parallax measurements safely reject membership in USCO.

2MASS~J04203904+2355502 was identified as a L1 candidate member of the TAU star-forming region by \cite{2006ApJ...645..676L}, but he noted that its with weak spectroscopic signatures of low surface gravity were indicative of an age between that of TAU and field brown dwarfs. Its trigonometric distance from \emph{Gaia}~DR2 ($38.7 \pm 1.3$\,pc) makes it a candidate member of the ABDMG in the foreground of TAU. A radial velocity measurement is still needed to fully confirm its membership in the ABDMG.

2MASS~J04254357+1616214 was identified by \citep{2008AA...481..661B} as a low-probability candidate member of the HYA, however its nearby distance ($67.0 \pm 0.3$\,pc) makes it a candiate member of the $\beta$PMG instead. The same is true of 2MASS~J04341301+1510569 \citep{1966AJ.....71..482V}, but its distance ($49.1 \pm 0.2$\,pc) makes it a viable candidate member of the ABDMG.

\begin{figure*}
	\centering
	\subfigure[$\beta$PMG]{\includegraphics[width=0.465\textwidth]{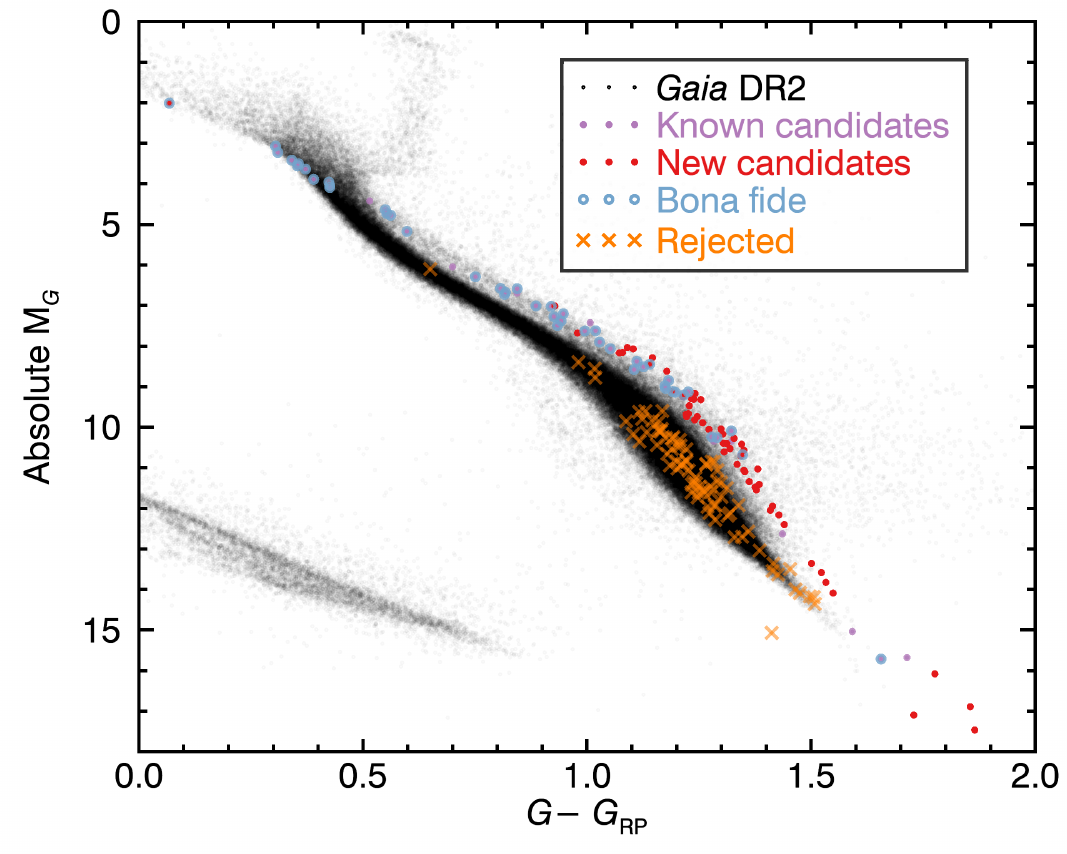}}
	\subfigure[THA, COL and CAR]{\includegraphics[width=0.465\textwidth]{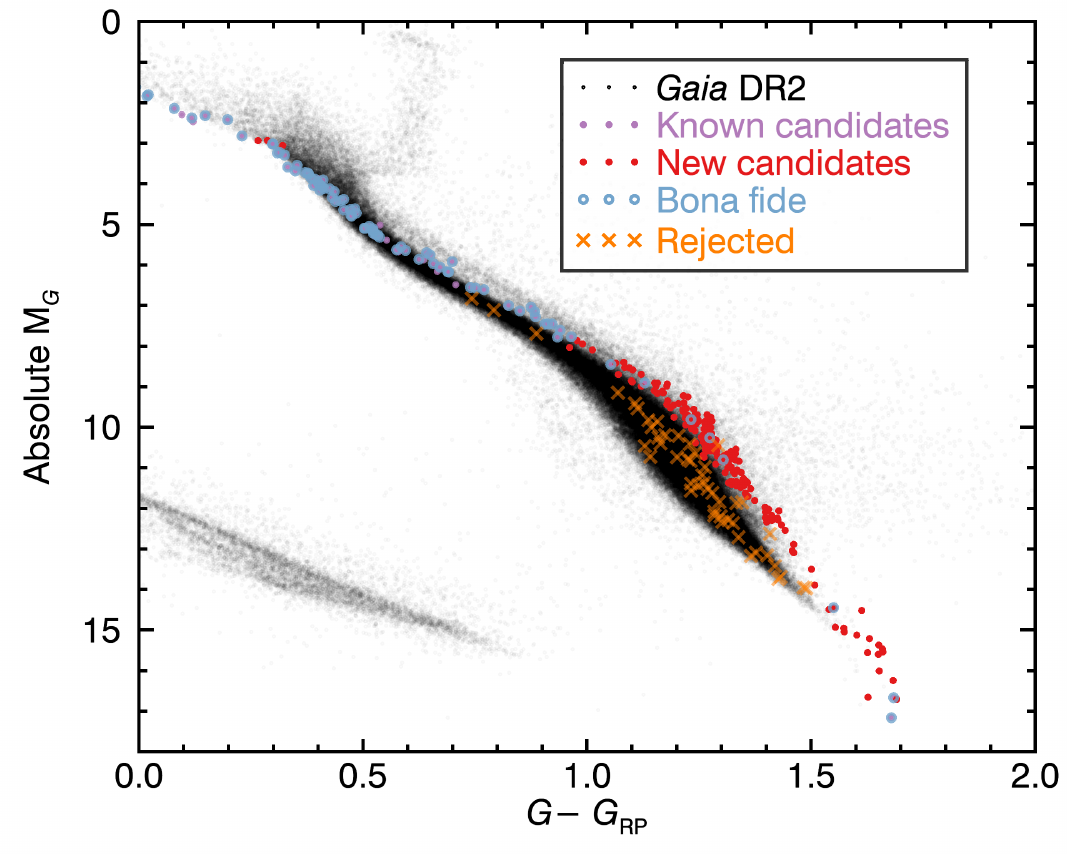}}
	\caption{Color-magnitude diagrams of known bona fide members (blue circles) and candidate members (filled purple circles) of young associations of similar ages \citep{2018ApJ...856...23G} compared to the new candidate members identified in this work (filled red circles). All candidates too faint to be consistent with their respective association (orange crosses) were rejected from our sample. See Section~\ref{sec:lit} for more detail.}
	\label{fig:cmdrej}
\end{figure*}

\added{We used \emph{Gaia} absolute $G$-band magnitude versus $G - G_{RP}$ color-magnitude diagrams to further reject any candidates not consistent with the known young members and candidate members or their respective young associations. The list of members compiled by \cite{2018ApJ...856...23G} were compared to our candidates one group at a time (e.g., see Figure~\ref{fig:cmdrej}, and any candidates significantly fainter than the young sequences were rejected from our sample. This step rejected an additional \rejectedbycmd\ candidates, most of them in the ABDMG (114), $\beta$PMG (79) and COL (37). This is consistent with the determination of \cite{2018ApJ...856...23G} that these associations are the most subject to contamination by unrelated field interlopers in kinematic-based searches, mostly because of their proximity (i.e., their members cover a larger fraction of the sky).}

\section{DISCUSSION}\label{sec:discussion}

\begin{figure}
	\centering
	\includegraphics[width=0.465\textwidth]{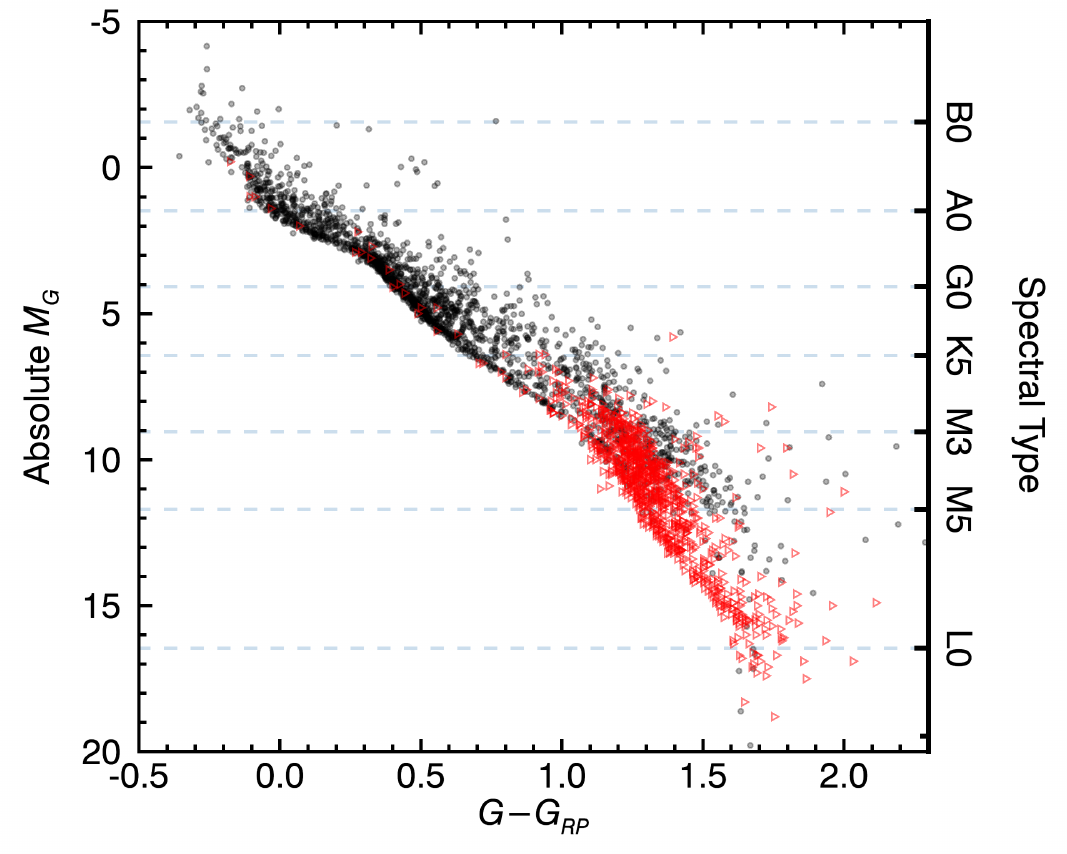}
	\caption{\emph{Gaia}~DR2 color-magnitude diagram of known bona fide members in young associations (black unfilled circles) compiled in \cite{2018ApJ...856...23G}. We choose $G-G_{RP}$ as a comparison color because photometric quality in \emph{Gaia}~DR2 remains good down to the substellar regime. New candidate members identified in this work are displayed as red unfilled circles. These data include pre-main sequence as well as main-sequence objects, depending on their masses and ages, which explains the large scatter especially at low masses (red colors). Some of the slightly older and more massive objects have also started to depart from the main sequence. See Section~\ref{sec:discussion} for more detail.}
	\label{fig:cmd_newcand}
\end{figure}

Our sample of new candidate members is the first step towards filling a gap in the current census of low-mass stars in young associations. Until now, obtaining trigonometric parallax measurements for a large number of low-mass stars has prevented completing the census of young associations especially in the M spectral class, which are fainter and much more numerous than the more massive members. Previous works have been successful at identifying a large fraction of the early M-type members (e.g., \citealp{2012ApJ...758...56S,2013ApJ...762...88M,2014ApJ...788...81M,2014AJ....147..146K,2017AJ....154...69S,Gagne:2018un}), but there remains a dearth of late-M type members. In Figure~\ref{fig:cmd_newcand}, we show a color-magnitude diagram of the current census of bona fide members \citep{2018ApJ...856...23G}, compared with the sample discussed here. This figure demonstrates how \emph{Gaia}~DR2 is particularly powerful at completing the faint, low-mass end of the color-magnitude diagram.

\begin{figure*}
	\centering
	\includegraphics[width=0.965\textwidth]{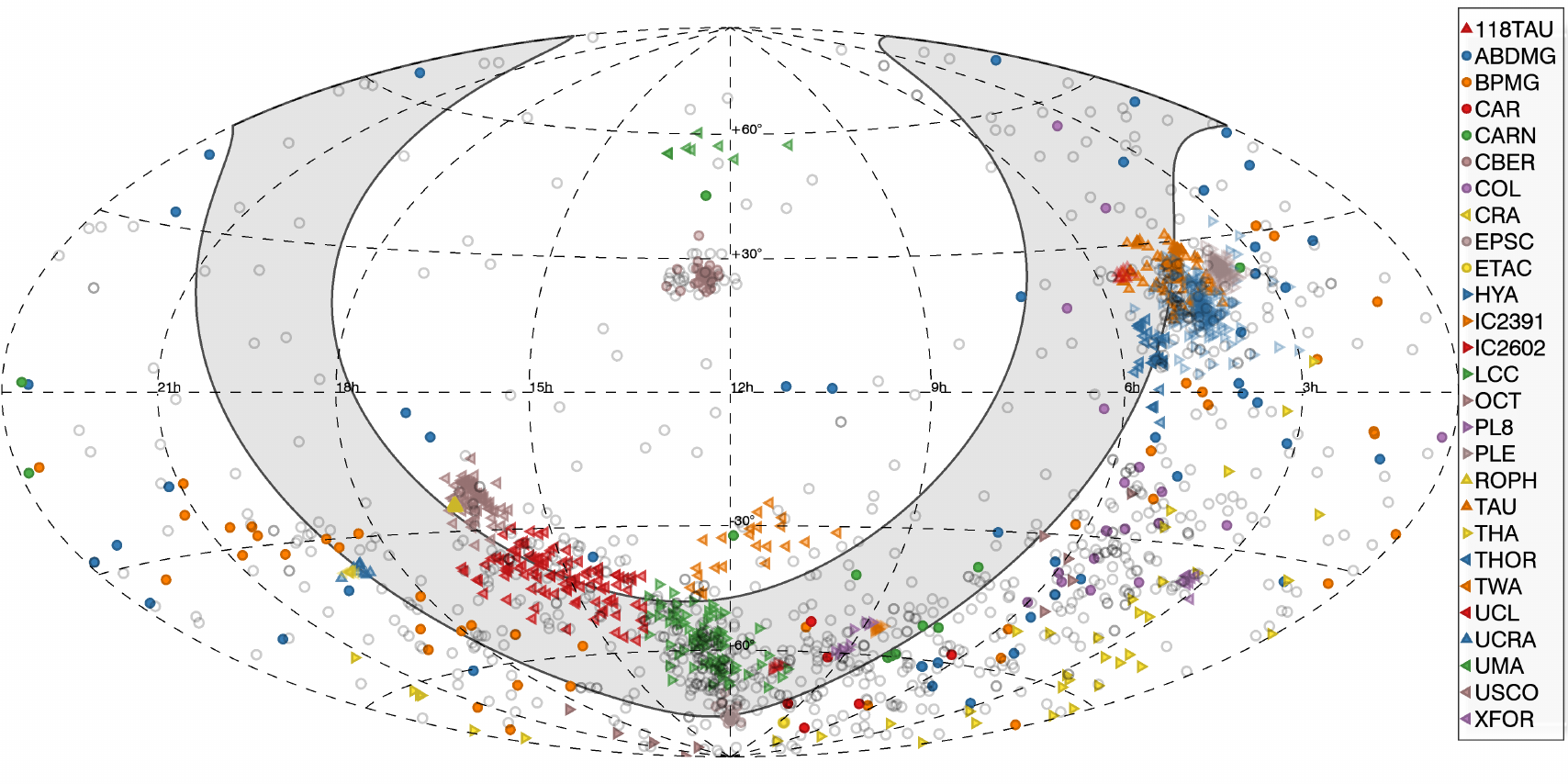}
	\caption{Sky position of known bona fide members in young associations (colored symbols, see legend for individual association names) compared with new candidate members uncovered here (black unfilled circles). The Galactic plane (Galactic latitudes $\pm$\,15\textdegree) is marked with a thick black line and a gray shaded region. The distribution of newly identified young associations generally tracks that of known members, and several new candidates were uncovered near the Galactic plane where previous searches were incomplete. See Section~\ref{sec:discussion} for more detail.}
	\label{fig:radec_newcand}
\end{figure*}

An advantage of \emph{Gaia}~DR2--based searches is that trigonometric distances make it possible to cover the Galactic plane with significantly less contamination than searches based on only proper motion. In Figure~\ref{fig:radec_newcand}, we show the distribution in sky positions of the new candidate members identified here. This figure demonstrates how our search is not biased away from the Galactic plane like most previous all-sky searches (e.g., \citealt{2015ApJ...798...73G}). Confirming membership still requires a spectroscopic follow-up to assess their youth and measure their radial velocities. In this section, we estimate the basic properties of the new candidates uncovered here in order to guide future telescope observations.


\begin{figure}
	\centering
	\includegraphics[width=0.465\textwidth]{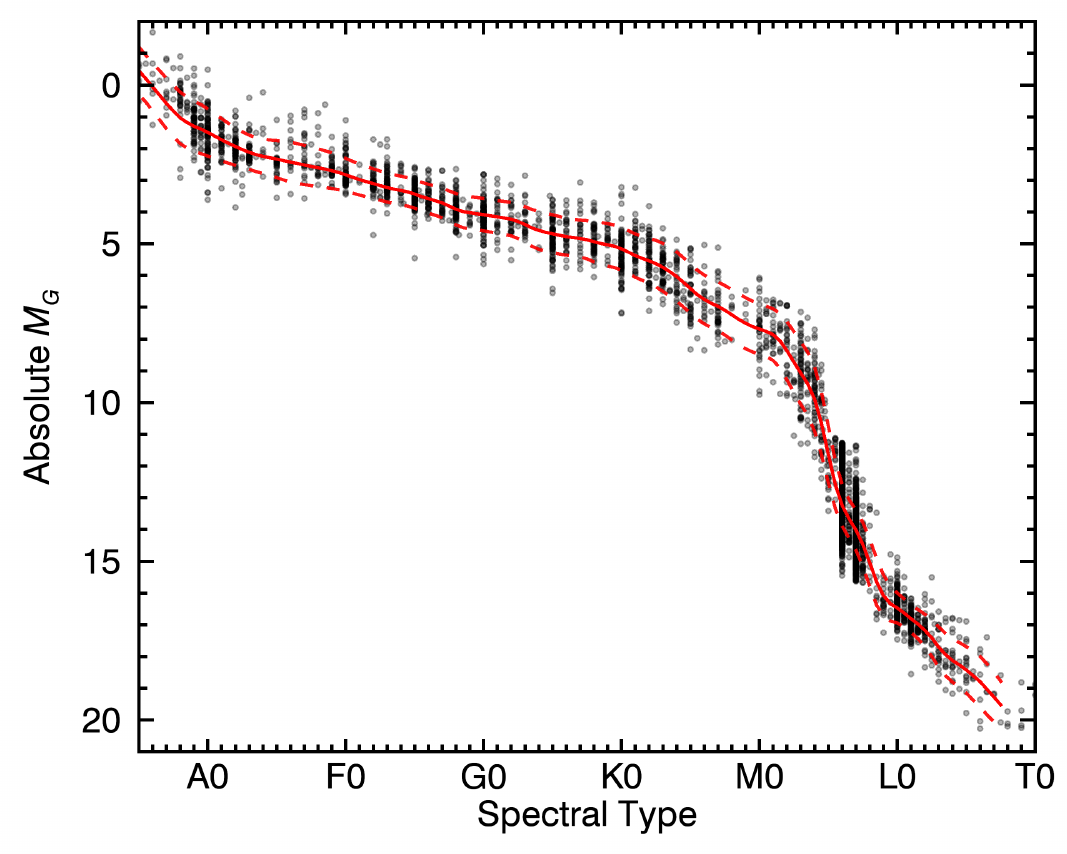}
	\caption{Absolute \emph{Gaia} $G$--band magnitude as a function of spectral type for various young and field objects (black circles). The median sequence is displayed as a solid red line and its standard deviation as dashed red lines. This median sequence was used to estimate the spectral types of new candidate members uncovered here, based on their absolute $G$--band magnitudes. Section~\ref{sec:spt} for more detail.}
	\label{fig:spt_mg_newcand}
\end{figure}
\begin{figure}
	\centering
	\includegraphics[width=0.465\textwidth]{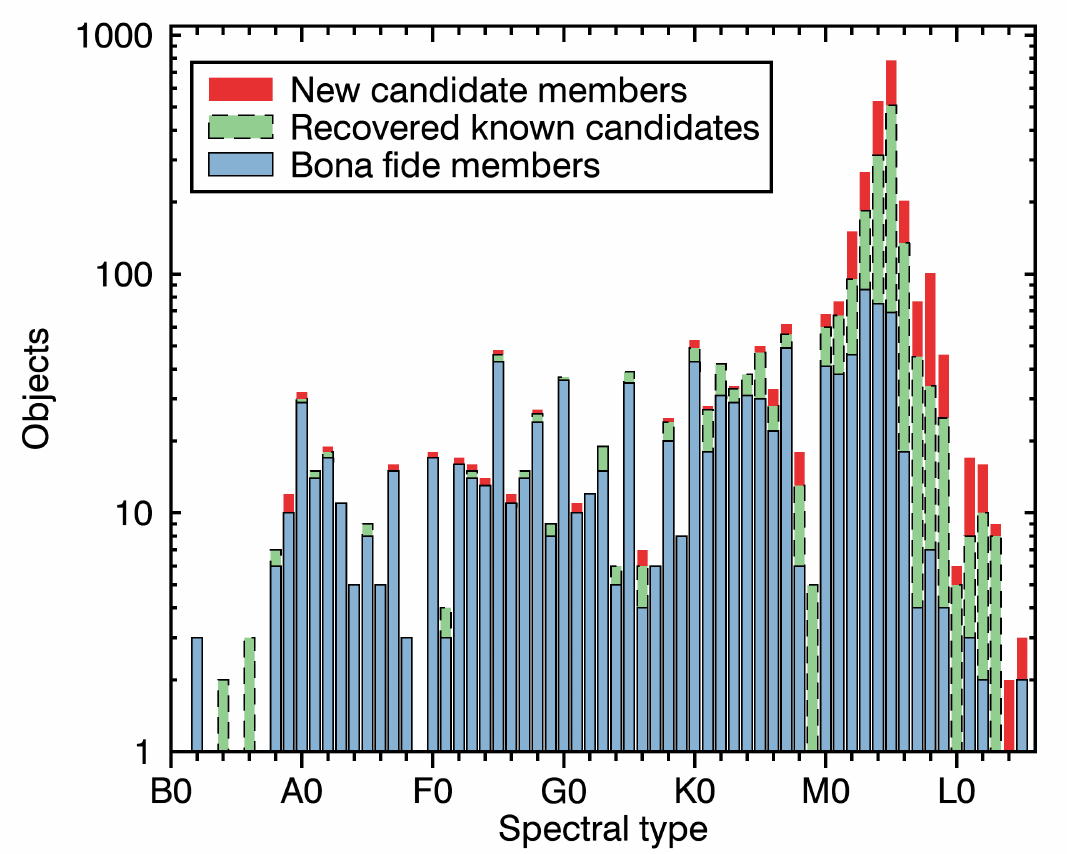}
	\caption{Spectral type histogram of known bona fide members of young associations (blue solid line), recovered known candidate members (green dashed lines), and new candidate members identfied in this work (red bars without outline). The contribution of this work is most important in the range of spectral types K7--L2.  See Section~\ref{sec:spt} for more detail.}
	\label{fig:spthist_newcand}
\end{figure}

\subsection{Photometric Spectral Type Estimates}\label{sec:spt}

In this section, we build an absolute $G$--band magnitude versus spectral type sequence in order to estimate the spectral types of the new young association candidates discovered here. We preferentially include young stars and brown dwarfs with ages $\sim$\,5--200\,Myr which are more representative of our sample, but we do not build separate sequences at different ages because only approximate spectral types will be calculated.

We cross-matched the lists of bona fide members of young associations compiled by \cite{2018ApJ...856...23G} with \emph{Gaia}~DR2 to build the bulk of the sequence across the stellar domain. Because the current list of bona fide members is incomplete for spectral types later than $\approx$\,K0, we added the lists of candidate members compiled by \cite{Gagne:2018un} and \cite{2018arXiv180405248G}, which cover spectral types down to $\approx$\,M5, and the known young brown dwarfs compiled by \cite{2016ApJS..225...10F} and \cite{2015ApJS..219...33G}. The resulting list of young objects contains a small number of objects in the ranges of spectral types M6--M9 and L5--T0, and we therefore completed these spectral ranges with the compilations of \cite{2002AJ....123.3409H,2007AJ....133..439C,2008AJ....135..785W} and \cite{2017MNRAS.469..401S}. These data likely do not represent the full spread in absolute $G$--band magnitudes that are due to young ages, but they remain useful to estimate spectral type.

We calculated the median absolute $G$--band magnitude and the corresponding standard deviation for each spectral type in the range B5--L9 with bins of 0.5 subtypes using the compilation described above. A 3$\sigma$ outlier rejection step was then applied, and the median and standard deviation across the full sequence was re-calculated. The resulting sequence is displayed in Figure~\ref{fig:spt_mg_newcand}, and was used to estimate the spectral types of the new candidates identified here unless they already had a measured spectral type in the literature. The average standard deviation around the resulting sequence is $\approx$\,0.6\,mag, which corresponds to a spectral type uncertainty of 3--5 subtypes, depending on spectral class. The resulting spectral type estimates are listed in Table~\ref{tab:candidates} with our compilation of new candidate members. We compare the distribution of estimated spectral types in the current sample with the known members and candidate members of young associations in Figure~\ref{fig:spthist_newcand}.

\begin{figure}
	\centering
	\includegraphics[width=0.465\textwidth]{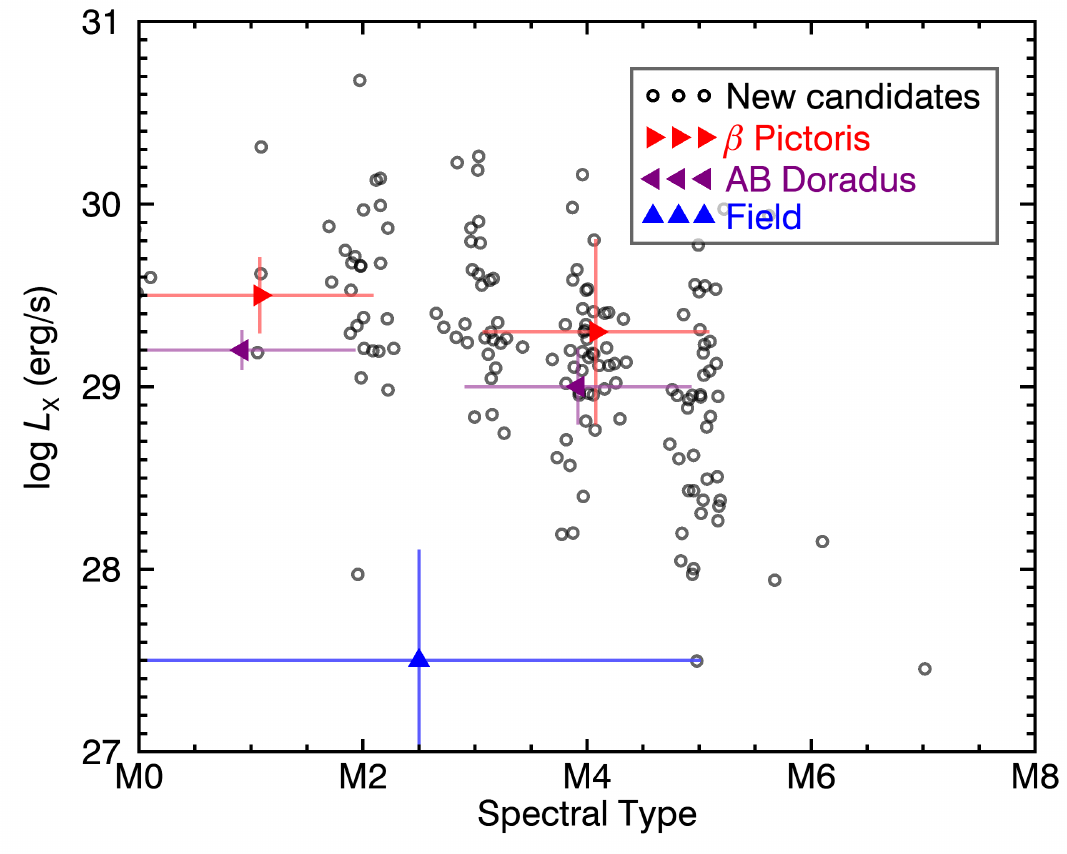}
	\caption{X-ray luminosity of our sample of new candidates (black unfilled circles) compared to the different samples of \cite{2014ApJ...788...81M}. Artifical noise with a standard deviation of 0.15 subtypes was added to our candidates for visibility. Most of the new candidates identified here have an X-ray luminosity consistent with the age of ABDMG or younger, as expected. See Section~\ref{sec:activ} for more detail.}
	\label{fig:xray}
\end{figure}
\begin{figure}
	\centering
	\includegraphics[width=0.465\textwidth]{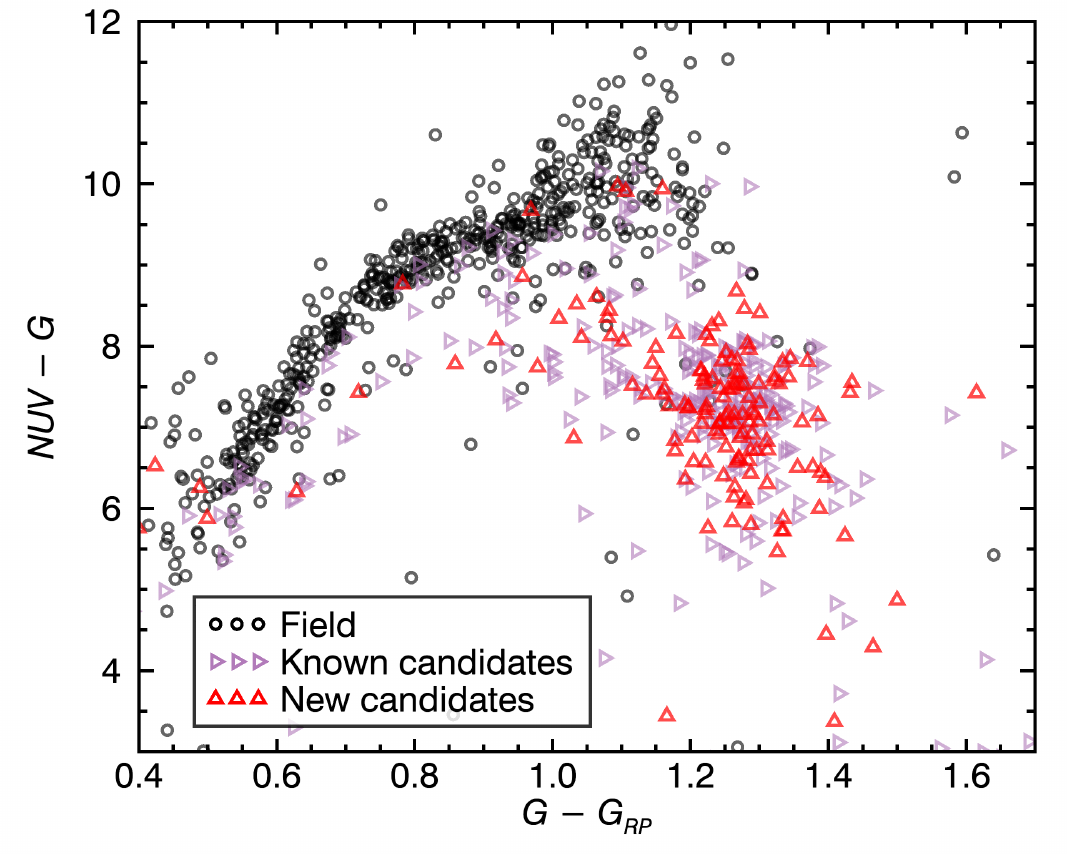}
	\caption{\emph{Galex} $NUV-G$ color versus \emph{Gaia} $G-R$ color for our sample of new candidates (red upward triangles) compared to known candidates and bona fide members recovered in our sample (rightward purple triangles) and field objects (black circles). Our sample mostly consists of highly active mid-M dwarfs, as would be expected at given the young ages of the associations considered here. See Section~\ref{sec:activ} for more detail.}
	\label{fig:galex}
\end{figure}

\subsection{Chromospheric Activity}\label{sec:activ}

One of the well known signs of youth for early- and mid-type M dwarfs is their chromospheric activity compared (e.g., \citealp{2011ApJ...727....6S,2011ApJ...727...62R,2014ApJ...788...81M,2015AJ....149..158S}). Signatures of high chromospheric activity include overluminosity at X-ray and ultraviolet wavelengths, as well as strong H$\alpha$ emission (e.g., \citealp{2008AJ....135..785W,2015AJ....149..158S}) and a high rate of flares (e.g., \citealp{2007AJ....133.2258S,2012ApJ...748...58D}). While assessing the latter two characteristics in our sample will require follow-up at the telescope in spectroscopy and imaging, the X-ray and ultraviolet properties of several new candidates presented here can already be obtained from the \emph{ROSAT} all-sky survey \citep{2016AA...588A.103B} and the \emph{GALEX} catalog \citep{2005ApJ...619L...1M}.

We cross-matched our sample of candidates with the second \emph{ROSAT} all-sky catalog to obtain their X-ray luminosity and compare them with the young and field samples of early- to mid-M dwarfs of \cite{2014ApJ...788...81M}. We used the \emph{Gaia}~DR2 proper motions and positions to project back positions to epoch 1994.5 and used a cross-match radius of 30$''$, yielding 194 matches\footnote{\added{30$"$ corresponds to the astrometric precision of ROSAT due to its large point-spread function. Contamination by background X-ray bright galaxies is expected to be very small given their small density on the sky (e.g., see \citealt{2004AA...425..367B}).}}. The resulting X-ray luminosities are compared with the field objects, and the ABDMG and $\beta$PMG candidate members of \cite{2014ApJ...788...81M} in Figure~\ref{fig:xray}. Most of our candidates demonstrate a clear X-ray over-luminosity compared to the field sample of \cite{2014ApJ...788...81M}, and are consistent with ages younger than $\sim$\,200\,Myr, which is expected based on the ages of the young associations considered here (see Table~\ref{tab:nyas}).

A similar cross-match was performed with the data release 5 of \emph{GALEX} (at epoch 2007.5 and a cross-match radius of 10$''$) and yielded 180 matches. The $NUV-G$ \emph{GALEX}--\emph{Gaia} color of our sample versus its $G-R$ \emph{Gaia} color (which traces spectral type) are compared to a sample of field stars and recovered known candidates or bona fide members in Figure~\ref{fig:galex}. Most of the new candidate members identified here have unusually blue $NUV-G$ colors compared to field stars, which is a signature of high chromospheric activity.

\subsection{Co-Moving Systems}\label{sec:comoving}

We searched the \emph{Gaia}~DR2 catalog for co-moving systems by examining a 2$'$ radius around each source in our sample of new candidate members. We searched for objects with proper motions within 10\,\masyr\ and a trigonometric parallax within 5\,mas of each other. We chose these very conservative limits to identify only the most likely co-moving systems while minimizing contamination from chance alignments. Deriving false-contamination probabilities at larger spatial or kinematic separations will require a careful examination of chance alignments within the members of a young association (e.g., see \citealt{2017ApJS..228...18G}).

Our conservative criteria identified \comovingsystems\ co-moving systems, 82 of which are co-moving with an object outside of our sample. Sixteen of the latter cases are new companions to known bona fide or candidate members of young associations. In \comovingrvs\ cases, we identifed a co-moving system that had a radial velocity measurement in the literature. Including this measurement in BANYAN~$\Sigma$ allowed us to reject \rejectedbycomovingrv\ candidate members from our sample and strengthen the membership of \strengthenedbycomovingrv\ others. All co-moving systems identified here are listed in Table~\ref{tab:comoving}.

\startlongtable
%\tabletypesize{\scriptsize}
% %\tablewidth{0.985\textwidth}
% %\setlength{\tabcolsep}{3pt}
 \begin{deluxetable}{lccc}
\tablecolumns{4}
\tablecaption{New likely L-type candidates identified in this work.\label{tab:bds}}
 \tablehead{\colhead{Spectral} & \colhead{Assoc.} & \colhead{R.A.\tablenotemark{b}} & \colhead{Decl.\tablenotemark{b}}\\
 \colhead{Type\tablenotemark{a}} & \colhead{} & \colhead{(hh:mm:ss.sss)} & \colhead{(dd:mm:ss.ss)}
}
\startdata
(L0)	&	ABDMG	&	01:38:47.53	&	-34:52:32.8	\\
(L1)	&	HYA	&	04:33:56.71	&	+05:37:23.7	\\
(L1)	&	HYA	&	04:45:43.83	&	+12:46:31.2	\\
(L2)	&	ABDMG	&	05:08:16.60	&	-14:13:49.6	\\
(L2)	&	CARN	&	05:19:28.80	&	-45:06:38.1	\\
(L1)	&	ABDMG	&	05:44:57.42	&	+37:05:00.5	\\
(L1)	&	COL	&	06:16:56.25	&	-25:43:55.9	\\
(L5)	&	CARN	&	07:23:52.61	&	-33:09:43.0	\\
(L1)	&	CAR	&	09:18:56.29	&	-61:01:18.7	\\
L2	&	CARN	&	09:28:39.53	&	-16:03:12.4	\\
(L2)	&	CARN	&	09:42:32.37	&	-25:51:37.6	\\
(L1)	&	CARN	&	11:50:42.86	&	-29:14:48.9	\\
L4	&	CARN	&	12:13:02.96	&	-04:32:44.3	\\
(L2)	&	BPMG	&	18:10:35.72	&	-55:13:45.0	\\
(L1)	&	BPMG	&	18:26:46.80	&	-46:02:24.5	\\
(L1)	&	ABDMG	&	18:40:59.39	&	-09:59:17.5	\\
(L3)	&	BPMG	&	21:04:31.36	&	-09:39:22.7	\\
(L1)	&	ABDMG	&	21:13:41.87	&	+35:07:39.6	\\
(L2)	&	ABDMG	&	23:54:12.69	&	+22:08:21.2	\\
\enddata
\tablenotetext{a}{Spectral types between parentheses were estimated from the absolute \emph{Gaia} $G$--band magnitude.}
\tablenotetext{b}{J2000 position at epoch 2015 from the \emph{Gaia}~DR2 catalog.}
\end{deluxetable}

\subsection{Isochrone Ages}\label{sec:isoc}

\begin{figure}
	\centering
	\includegraphics[width=0.465\textwidth]{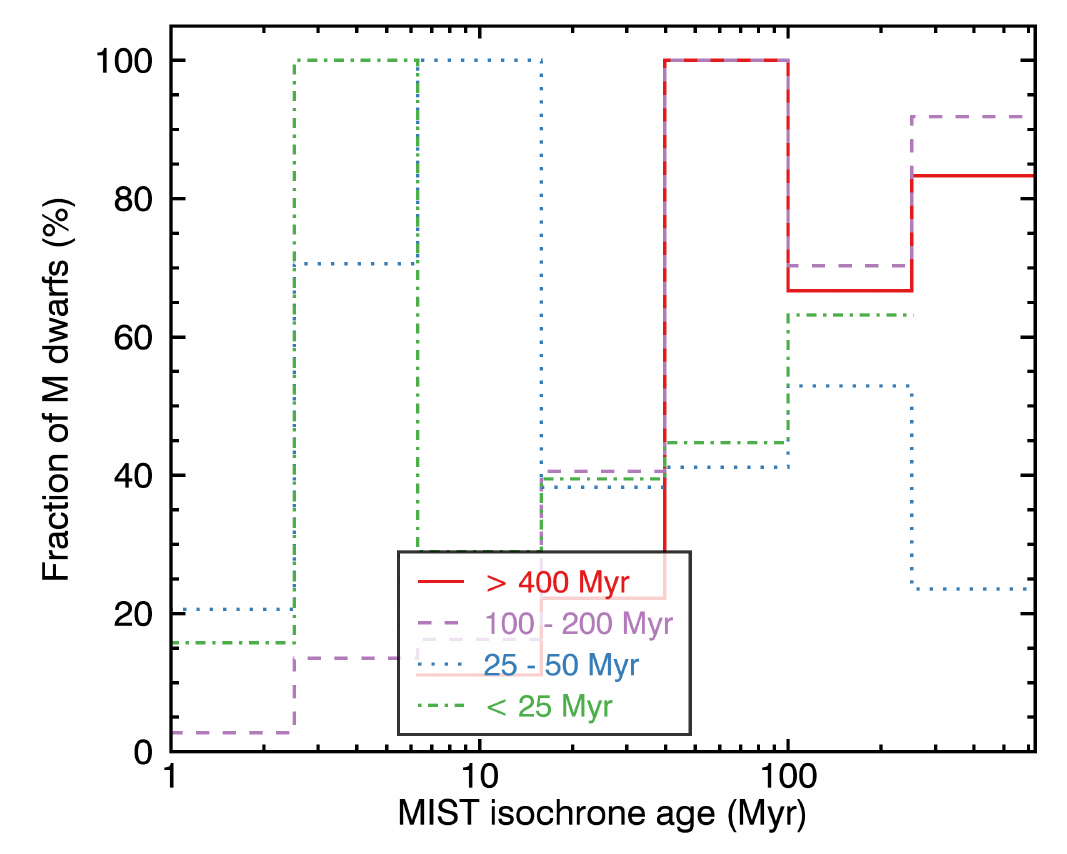}
	\caption{Distribution of MIST isochrone ages for new M-type candidate members in young associations of different age ranges. These age estimates ignore the effects of magnetic fields and are therefore expected to be systematically too low, but the relative differences in age distribution between candidates in associations of different ages can already be seen clearly. See Section~\ref{sec:isoc} for more detail.}
	\label{fig:isoc}
\end{figure}

We compared the absolute \emph{Gaia} $G$-band magnitudes versus $G-R$ color of our new mid to late M-type candidates ($G - G_{RP} > 1.3$\,mag) with those of the MIST solar-metallicity tracks v1.1 \citep{2016ApJ...823..102C}. We used the Bayesian method described by \cite{Gagne:2018un} to verify if we find relative differences in the distribution of isochrone ages in our candidate members of associations with different ages. The resulting isochrone age distributions for our targets are displayed in Figure~\ref{fig:isoc}. We note that the MIST isochrone age estimations ignore the contribution of magnetic fields. This is known to produce ages that are systematically too young compared to other age-dating methods such as the lithium depletion boundary for M dwarfs \citep{2014ApJ...792...37M}. However, we can clearly see a correlation between the distributions of isochrone ages and the expected ages of the candidates based on their respective young associations. These new M dwarf candidates will be valuable to better understand the distribution of magnetic field strength versus age after a detailed spectrosopic follow-up is completed.

\begin{figure}
	\centering
	\includegraphics[width=0.465\textwidth]{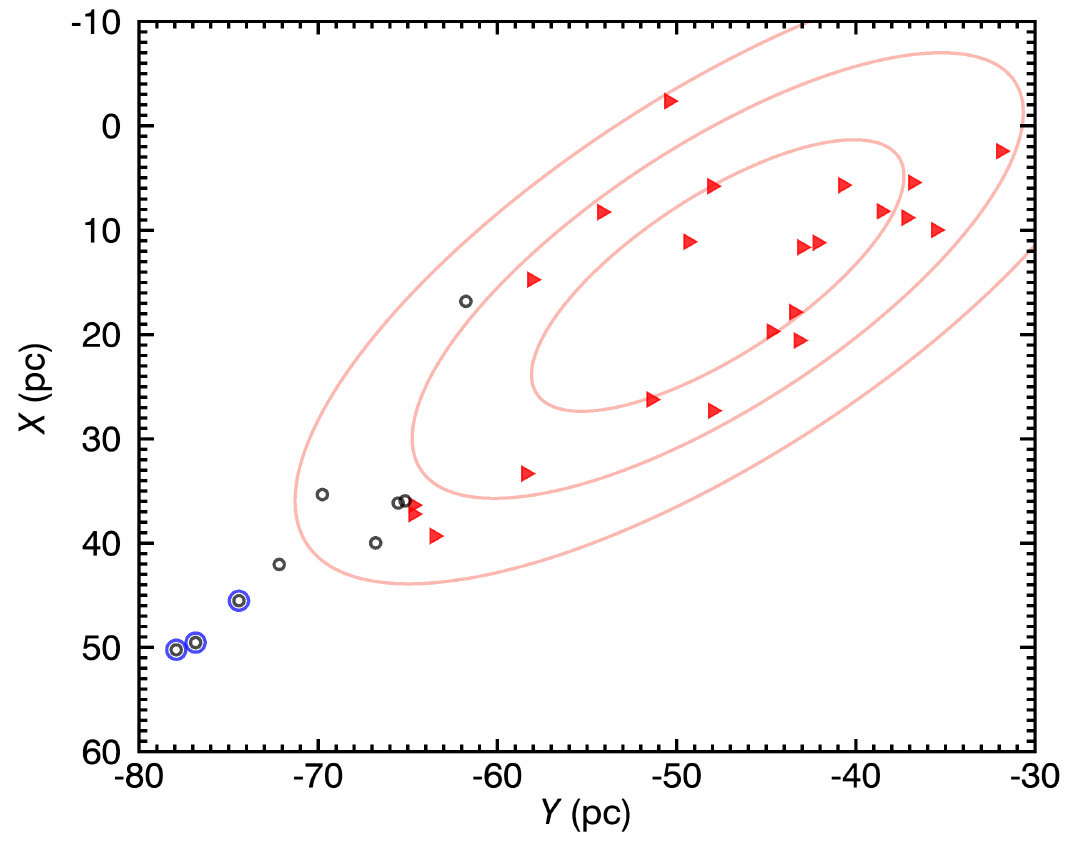}
	\caption{Galactic positions $X$ and $Y$ of known TWA members (rightward red triangles) compared to new candidates identified in this work (black unfilled circles). The new candidates identified here form an extension of the TWA distribution to slightly larger distances. Three of the new TWA candidates have a non-negligible membership probability in LCC (marked with larger blue unfilled circles) and may form a bridge between the two associations. The 1, 2 and 3$\sigma$ contours of the BANYAN~$\Sigma$ spatial model of TWA are shown with thick orange lines. See Section~\ref{sec:ext} for more detail.}
	\label{fig:twa_ext}
\end{figure}

\begin{figure*}
	\centering
	\subfigure[CAR]{\includegraphics[width=0.488\textwidth]{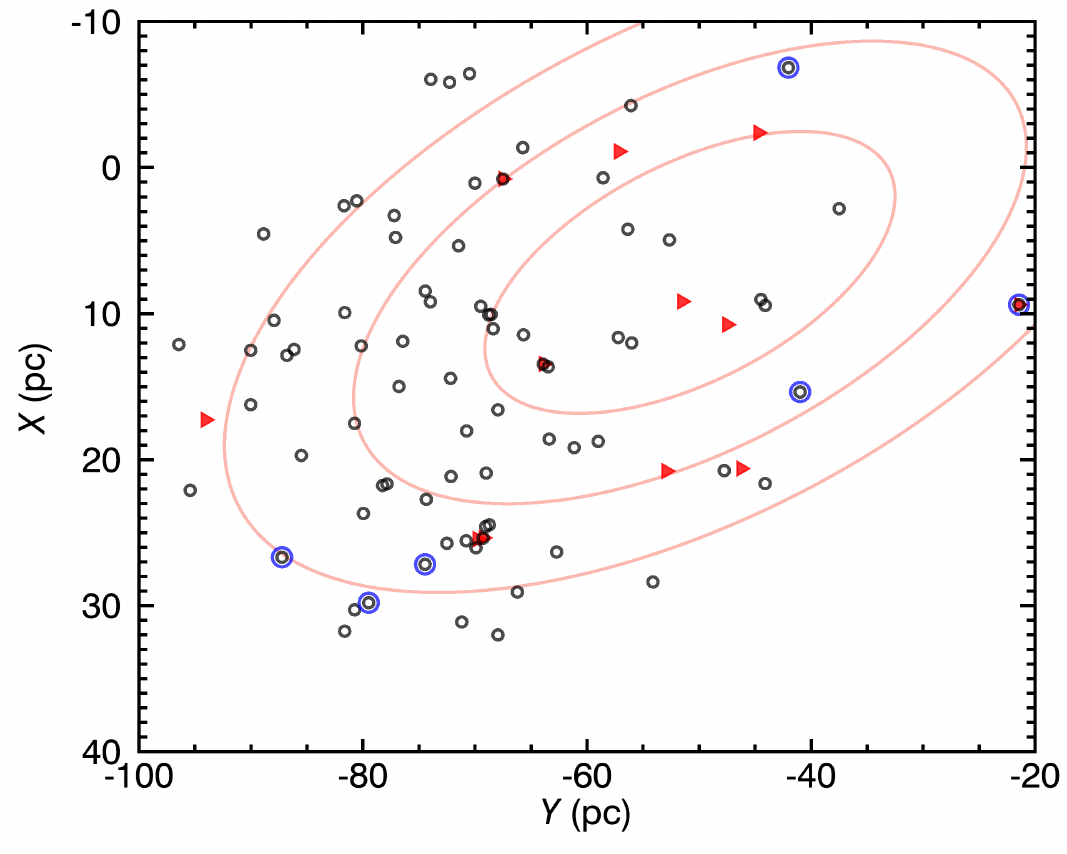}}
	\subfigure[COL]{\includegraphics[width=0.488\textwidth]{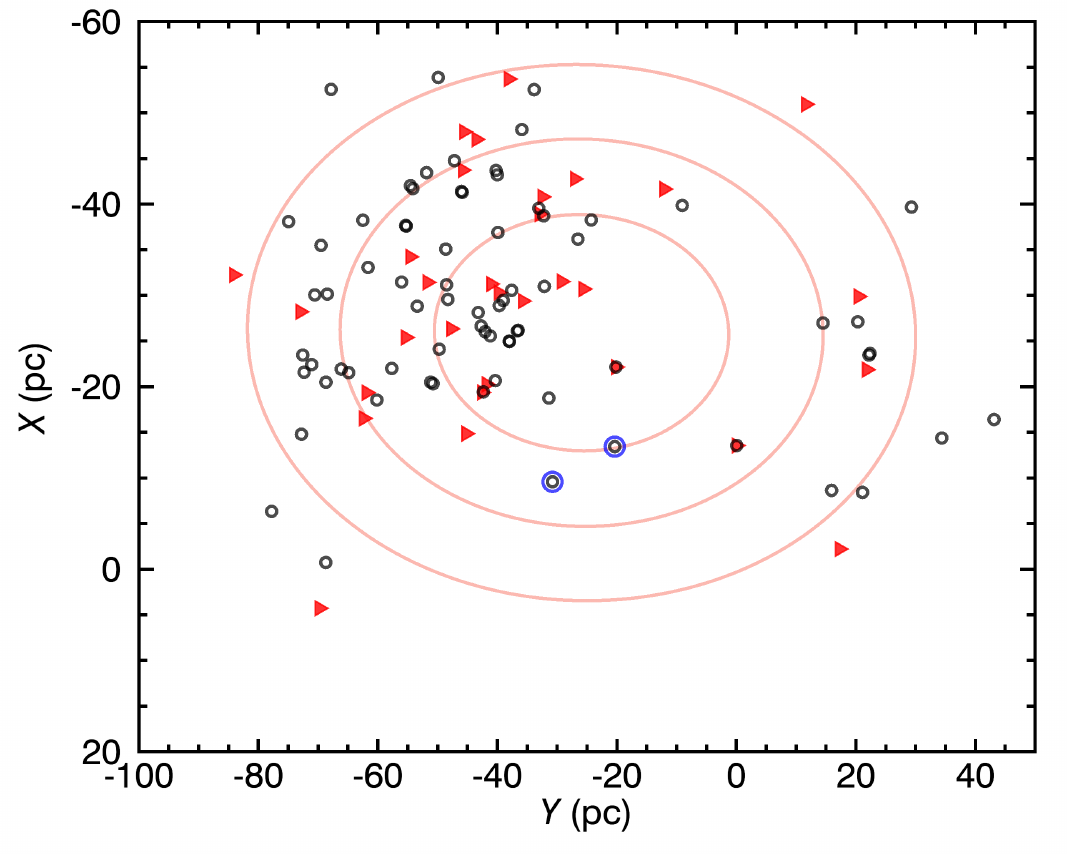}}
	\caption{Galactic positions $X$, $Y$ and $Z$ of known members in CAR and COL (rightward red triangles) compared to new candidates identified in this work (black unfilled circles). The 1, 2 and 3$\sigma$ contours of the BANYAN~$\Sigma$ spatial models of the respective groups are shown with thick orange lines. The candidates with a non-negligible membership probability in more than one young association are marked with larger blue unfilled circles. The new candidate members identified in this survey show slightly different spatial distributions than the known members. This could be caused by non-uniform contamination, a bias in the inferred distribution caused by the small number of known bona fide members, or a mass dependency on the spatial distribution of members. The distance limit of 100\,pc that we imposed on our sample did not affect these nearby objects. See Section~\ref{sec:ext} for more detail.}
	\label{fig:XYZ}
\end{figure*}

\subsection{Brown Dwarfs}\label{sec:bds}

We used the bolometric luminosity versus spectral type relations of \cite{2016ApJS..225...10F} combined with the \cite{2008ApJ...689.1327S} evolutionary tracks to obtain a rough estimate of the spectral types that correspond to the substellar regime at the age of each young association considered here. We used these estimates to identify a total of \browndwarfs\ objects in our sample that are new brown dwarf candidate members in young associations with spectral types later than M6--L2, depending on the age of the young association they likely belong to. Obtaining near-infrared spectra yield gravity-sensitive spectroscopic indices (e.g., \citealt{2013ApJ...772...79A}), and will provide valuable data for interpreting atmospheres similar to those of gaseous giant planets (e.g., \citealp{2008Sci...322.1348M,2013ApJ...772L..15R}). Ultimately this sample will place constraints on the very low-mass end of the initial mass function (e.g., \citealt{2015ApJS..219...33G}). There are \ldwarfs\ brown dwarfs in our sample with estimated spectral types in the L class.\added{ Their parallax measurements are 7--90 times larger than the measurement errors, with a median or 25}. They are listed in Table~\ref{tab:bds}.

\subsection{Tentative Indications of Spatial Extension}\label{sec:ext}

It can be expected that the current census of the nearby young associations is biased toward nearby distances, because our knowledge of their spatial distribution is still based on members bright enough to have been detected by the Hipparcos mission \citep{1997AA...323L..49P}. Recent discoveries \citep{2017AJ....153...18B,2018ApJ...852...55D} suggested for example that the ABDMG might be larger than previously thought, and its very similar kinematics and age to the more distant Pleiades association (e.g., see \citealp{2005ApJ...628L..69L,2014prpl.conf..219S,2018ApJ...856...23G}) raises the question of whether the two associations may be related. In its default setting, BANYAN~$\Sigma$ compares the $XYZ$ Galactic positions of objects to a fixed spatial model density of young associations, and is therefore not designed to uncover such spatial extensions of known associations. Our search did not identify candidate members of ABDMG at larger distances, but this could be entirely caused by our bias in recovering new candidate members that are located spatially near those already known. Future work will be needed to explore this possibility. BANYAN~$\Sigma$ has the option to consider only $UVW$ space velocities in its search for new candidate members, but in this case it will be important to confirm the youth of the candidate members bceause searches based on $UVW$ only will suffer significant contamination from random field interlopers.

Our survey has however identified a possible extension of the TWA to slightly larger distances (see Figure~\ref{fig:twa_ext}) that may bridge it to the LCC region of the Sco-Cen star-formation complex. Searches for new candiate members in TWA have been known to be contaminated by the LCC region of the Sco-Cen star-formation complex (e.g., see \citealt{2017ApJS..228...18G}), however BANYAN~$\Sigma$ now includes a model of LCC. We uncovered 5 candidates at the far end of TWA that have a zero LCC membership probability, and 4 slightly more distant candidates that are most likely TWA candidates but have non-negligible membership probabilities in LCC. Determining the ages of these stars seemingly bridging the two associations may allow us to understand whether they are simply contaminants or whether TWA and LCC are related.

Our sample of new candidates in CAR and COL also show slightly different spatial distributions than the currently known bona fide members (see Figure~\ref{fig:XYZ}). Confirming the radial velocities and youth of the new candidates will be required to determine whether the spatial models of BANYAN~$\Sigma$ will need to be refined, and to test whether the spatial distribution of the lower-mass members are distinct from those of the more massive members. The distributions of new candidate members closely followed that of known bona fide members for all other associations and spatial-kinematic projections.

\subsection{New Bona Fide Members}\label{sec:newbf}

There are 17 B9--G8 stars in our sample that have full kinematics and were not previously recognized as candidate members of young associations in the literature. We calculated MIST isochrone ages for them (see Section \ref{sec:isoc} for more detail) and report them in Table~\ref{tab:bfs} as new bona fide members of their respective young associations. All of them have ages consistent with their young association, except for iot~Cen ($7^{-2}_{+4}$\,Myr) which is much too young to belong in CARN. The ages of the F-type stars in Table~\ref{tab:bfs} cannot be constrained with isochrones, and it will be difficult to pose any age constraints for these stars. In addition to these, there are 15 stars in our sample that are co-moving with a known bona fide member (see Section~\ref{sec:comoving}). Those are also reported in Table~\ref{tab:bfs}.

\section{CONCLUSIONS}\label{sec:conclusion}

We used the BANYAN~$\Sigma$ algorithm with the nearest 100\,pc entries in the \emph{Gaia}~DR2 to identify \finalcandidates\ new candidate members with spectral types in the range B9--L2 in the 27 nearest young associations from an initial sample of \initialobjects\ candidates (\knownpluskastnerobjects\ were already known in the literature, \rejectedbyvisual\ \emph{Gaia}~DR2 entries were rejected from a visual examination of finder charts, and \rejectedbyrv\ were rejected with a literature radial velocity measurement). \comovingsystems\ of these objects are co-moving systems, and will serve as important benchmarks for the calibration of atmosphere and evolutionary models. Spectroscopic follow-up of these targets to obtain radial velocities and signatures of youth such as lithium absorption will be needed to confirm their membership. Our search identified mostly mid-M candidate members, which contribute most importantly to the initial mass function. The ongoing TESS mission \citep{2015JATIS...1a4003R} will uncover exoplanet companions to several of these nearby M dwarfs with the transit method. Identifying age-calibrated stars among the TESS sample in 27 distinct populations with ages in the range $\sim$\,2--750\,Myr will provide crucial information on the fundamental parameters of these future exoplanet discoveries, and will inform how the structure of planetary systems evolve with time.

\acknowledgments

\added{We thank the anonymous referee for thoughtful comments that improved the quality of this work. }We thank Ricky L. Smart\added{ and Joel Kastner} for useful comments. This research made use of: the SIMBAD database and VizieR catalog access tool, operated at the Centre de Donn\'ees astronomiques de Strasbourg, France \citep{2000AAS..143...23O}; data products from the Two Micron All Sky Survey (\emph{2MASS}; \citealp{2006AJ....131.1163S}), which is a joint project of the University of Massachusetts and the Infrared Processing and Analysis Center (IPAC)/California Institute of Technology (Caltech), funded by the National Aeronautics and Space Administration (NASA) and the National Science Foundation \citep{2006AJ....131.1163S}; data products from the \emph{Wide-field Infrared Survey Explorer} (\emph{WISE}; and \citealp{2010AJ....140.1868W}), which is a joint project of the University of California, Los Angeles, and the Jet Propulsion Laboratory (JPL)/Caltech, funded by NASA. The Digitized Sky Surveys were produced at the Space Telescope Science Institute under U.S. Government grant NAG W-2166. The images of these surveys are based on photographic data obtained using the Oschin Schmidt Telescope on Palomar Mountain and the UK Schmidt Telescope. The plates were processed into the present compressed digital form with the permission of these institutions. The Second Palomar Observatory Sky Survey (POSS-II) was made by the California Institute of Technology with funds from the National Science Foundation, the National Geographic Society, the Sloan Foundation, the Samuel Oschin Foundation, and the Eastman Kodak Corporation. The Oschin Schmidt Telescope is operated by the California Institute of Technology and Palomar Observatory. This work presents results from the European Space Agency (ESA) space mission Gaia. Gaia data are being processed by the Gaia Data Processing and Analysis Consortium (DPAC). Funding for the DPAC is provided by national institutions, in particular the institutions participating in the Gaia MultiLateral Agreement (MLA). The Gaia mission website is https://www.cosmos.esa.int/gaia. The Gaia archive website is https://archives.esac.esa.int/gaia. This research was started at the NYC Gaia DR2 Workshop at the Center for Computational Astrophysics of the Flatiron Institute in 2018 April.

\startlongtable
\tabletypesize{\scriptsize}
% %\tablewidth{0.985\textwidth}
% %\setlength{\tabcolsep}{3pt}
\begin{longrotatetable}
\global\pdfpageattr\expandafter{\the\pdfpageattr/Rotate 90}
\begin{deluxetable*}{lcccccccccc}
\tablecolumns{11}
\tablecaption{New candidates identified in this work.\label{tab:candidates}}
 \tablehead{
 \colhead{Name} & \colhead{Spectral} & \colhead{Assoc.} & \colhead{R.A.\tablenotemark{b}} & \colhead{Decl.\tablenotemark{b}} & \colhead{$\mu_\alpha\cos\delta$} & \colhead{$\mu_\delta$} & \colhead{Parallax} & \colhead{Rad. Vel.} & \colhead{Ambiguous} & \colhead{Ref.\tablenotemark{d}}\\
 \colhead{} & \colhead{Type\tablenotemark{a}} & \colhead{} & \colhead{(hh:mm:ss.sss)} & \colhead{(dd:mm:ss.ss)} & \colhead{(\masyr)} & \colhead{(\masyr)} & \colhead{(mas)} & \colhead{(\kms)} & \colhead{Assoc.\tablenotemark{c}} & \colhead{}
}
\startdata
J0644-4137 & (M4) & ABDMG & 06:44:54.065 & -41:37:27.38 & $-5.17 \pm 0.07$ & $-20.89 \pm 0.07$ & $20.37 \pm 0.04$ & $\cdots$ & $\cdots$ & --,--\\
J0646+1015 & (M6) & ABDMG & 06:46:05.803 & +10:15:37.96 & $-19.38 \pm 0.38$ & $-98.25 \pm 0.44$ & $16.95 \pm 0.14$ & $\cdots$ & $\cdots$ & --,--\\
J0650-3203 & (M5) & ABDMG & 06:50:27.278 & -32:03:19.11 & $-26.22 \pm 0.08$ & $-54.64 \pm 0.09$ & $31.70 \pm 0.05$ & $\cdots$ & $\cdots$ & --,--\\
J0700-4447 & (M6) & ABDMG & 07:00:34.277 & -44:47:13.56 & $-43.50 \pm 0.25$ & $-14.42 \pm 0.25$ & $45.22 \pm 0.13$ & $\cdots$ & $\cdots$ & --,--\\
J0703-4152 & (M9) & ABDMG & 07:03:02.270 & -41:52:22.68 & $-11.7 \pm 1.2$ & $-20.2 \pm 2.5$ & $20.72 \pm 0.86$ & $\cdots$ & $\cdots$ & --,--\\
J0703-6110 & (M8) & ABDMG & 07:03:23.851 & -61:10:03.23 & $-11.27 \pm 0.68$ & $26.39 \pm 0.63$ & $15.94 \pm 0.32$ & $\cdots$ & $\cdots$ & --,--\\
J0703-1723 & (M4) & ABDMG & 07:03:38.676 & -17:23:39.06 & $-39.40 \pm 0.17$ & $-77.74 \pm 0.14$ & $24.67 \pm 0.10$ & $\cdots$ & $\cdots$ & --,--\\
J0703-5505 & (M8) & ABDMG & 07:03:59.359 & -55:05:58.52 & $-27.22 \pm 0.44$ & $11.64 \pm 0.42$ & $19.83 \pm 0.19$ & $\cdots$ & $\cdots$ & --,--\\
\enddata
\end{deluxetable*}
\end{longrotatetable}
\global\pdfpageattr\expandafter{\the\pdfpageattr/Rotate 0}
\startlongtable
\tabletypesize{\scriptsize}
%\tablewidth{0.985\textwidth}
%\setlength{\tabcolsep}{3pt}
\begin{deluxetable*}{lcccccccc}
\tablecolumns{9}
\tablecaption{Co-moving systems identified in this work.\label{tab:comoving}}
\tablehead{\colhead{\emph{Gaia}~DR2} & \colhead{Assoc.} & \colhead{Separation} & \colhead{R.A.\tablenotemark{a}} & \colhead{Decl.\tablenotemark{a}} & \colhead{$\mu_\alpha\cos\delta$} & \colhead{$\mu_\delta$} & \colhead{Parallax} & \colhead{Rad. Vel.}\\
\colhead{Source ID} & \colhead{} & \colhead{($''$)} & \colhead{(hh:mm:ss.sss)} & \colhead{(dd:mm:ss.ss)} & \colhead{(\masyr)} & \colhead{(\masyr)} & \colhead{(mas)} & \colhead{(\kms)}}
\startdata
5000558409016727296 & THA & $\cdots$ & 00:39:35.477 & -38:17:18.69 & $101.47 \pm 0.10$ & $-66.50 \pm 0.07$ & $24.92 \pm 0.06$ & $(2.7 \pm 0.7)$\\
5000558443376465408 & $\cdots$ & 19.9 & 00:39:35.930 & -38:16:59.55 & $99.45 \pm 0.08$ & $-64.46 \pm 0.07$ & $24.85 \pm 0.04$ & $1.7 \pm 4.0$\\
2371914419569514752 & ABDMG & $\cdots$ & 01:06:25.510 & -14:17:28.82 & $101.82 \pm 0.11$ & $-98.76 \pm 0.08$ & $22.02 \pm 0.05$ & $(9.5 \pm 1.8)$\\
2371914419569515136 & $\cdots$ & 22.6 & 01:06:26.260 & -14:17:48.57 & $99.93 \pm 0.08$ & $-94.63 \pm 0.06$ & $22.01 \pm 0.04$ & $8.0 \pm 0.3$\\
292521529517547776 & ABDMG & $\cdots$ & 01:39:58.869 & +26:11:00.80 & $138.98 \pm 0.26$ & $-146.21 \pm 0.12$ & $31.02 \pm 0.08$ & $(-4.0 \pm 1.4)$\\
292521529517332224 & $\cdots$ & 8.0 & 01:39:59.341 & +26:11:05.74 & $135.06 \pm 0.18$ & $-151.91 \pm 0.09$ & $30.85 \pm 0.05$ & $-8.9 \pm 10.1$\\
4941694920153720448 & ABDMG & $\cdots$ & 01:48:47.900 & -48:31:16.38 & $110.54 \pm 0.06$ & $-53.67 \pm 0.06$ & $25.62 \pm 0.04$ & $(21.7 \pm 1.2)$\\
4941706636824503680 & $\cdots$ & 72.0 & 01:48:41.060 & -48:30:52.75 & $110.90 \pm 0.04$ & $-53.97 \pm 0.04$ & $25.54 \pm 0.02$ & $20.9 \pm 1.3$\\
\enddata
\tablenotetext{a}{J2000 position at epoch 2015 from the \emph{Gaia}~DR2 catalog.}
\end{deluxetable*}

\startlongtable
%\tabletypesize{\scriptsize}
% %\tablewidth{0.985\textwidth}
% %\setlength{\tabcolsep}{3pt}
 \begin{deluxetable*}{llccccl}
\tablecolumns{7}
\tablecaption{New bona fide members.\label{tab:bfs}}
 \tablehead{\colhead{Name} & \colhead{Spectral} & \colhead{Assoc.} & \colhead{R.A.\tablenotemark{b}} & \colhead{Decl.\tablenotemark{b}} & \colhead{MIST} & \colhead{Co-moving}
 \\
 \colhead{} & \colhead{Type\tablenotemark{a}} & \colhead{} & \colhead{(hh:mm:ss.sss)} & \colhead{(dd:mm:ss.ss)} & \colhead{Age (Myr)} & \colhead{}
}
\startdata
\sidehead{\textbf{New Bona Fide Members}}													
gam02 Ari	&	A2IV	&	ABDMG	&	01:53:31.90	&	+19:17:36.4	&	4--300	&	$\cdots$	\\
gam01 Ari	&	A0V	&	ABDMG	&	01:53:31.90	&	+19:17:43.8	&	4--300	&	$\cdots$	\\
HD 14691	&	F3V	&	COL	&	02:22:01.69	&	-10:46:40.4	&	10--3000	&	$\cdots$	\\
iot Hyi	&	F5IV-V	&	CAR	&	03:15:58.20	&	-77:23:17.5	&	$>$ 10	&	$\cdots$	\\
ups04 Eri	&	B9V	&	THA	&	04:17:53.74	&	-33:47:54.1	&	1--200	&	$\cdots$	\\
HD 36169	&	G8V	&	ABDMG	&	05:29:03.29	&	-19:08:35.9	&	30--1000	&	$\cdots$	\\
HD 43199	&	F0III/IV	&	CAR	&	06:10:52.92	&	-61:29:58.8	&	$\cdots$	&	$\cdots$	\\
HD 80950	&	A0V	&	CAR	&	09:17:27.45	&	-74:44:04.0	&	5--900	&	$\cdots$	\\
\enddata
\tablenotetext{a}{Spectral types between parentheses were estimated from the absolute \emph{Gaia} $G$--band magnitude.}
\tablenotetext{b}{J2000 position at epoch 2015 from the \emph{Gaia}~DR2 catalog.}
\tablenotetext{c}{The full name of this object is 2MASS~J00393579--3816584.}
\tablenotetext{d}{The full name of this object is 2MASS~J16572029--5343316.}
\end{deluxetable*}

\emph{JG} wrote the codes, manuscript, generated figures and led all analysis; \emph{JKF} provided help with parsing the \emph{Gaia}~DR2 data as well as general comments.

\software{BANYAN~$\Sigma$ \citep{2018ApJ...856...23G}.}

\bibliographystyle{apj}

\begin{thebibliography}{}
\expandafter\ifx\csname natexlab\endcsname\relax\def\natexlab#1{#1}\fi

\bibitem[{Abt \& Cardona(1984)}]{1984ApJ...276..266A}
Abt, H.~A., \& Cardona, O. 1984, The Astrophysical Journal, 276, 266

\bibitem[{Abt \& Morrell(1995)}]{1995ApJS...99..135A}
Abt, H.~A., \& Morrell, N.~I. 1995, Astrophysical Journal Supplement v.99, 99,
  135

\bibitem[{Alam {et~al.}(2015)Alam, Albareti, Allende~Prieto, Anders, Anderson,
  Anderton, Andrews, Armengaud, Aubourg, Bailey, Basu, Bautista, Beaton, Beers,
  Bender, Berlind, Beutler, Bhardwaj, Bird, Bizyaev, Blake, Blanton, Blomqvist,
  Bochanski, Bolton, Bovy, Shelden~Bradley, Brandt, Brauer, Brinkmann, Brown,
  Brownstein, Burden, Burtin, Busca, Cai, Capozzi, Carnero~Rosell, Carr,
  Carrera, Chambers, Chaplin, Chen, Chiappini, Chojnowski, Chuang, Clerc,
  Comparat, Covey, Croft, Cuesta, Cunha, da~Costa, Da~Rio, Davenport, Dawson,
  De~Lee, Delubac, Deshpande, Dhital, Dutra-Ferreira, Dwelly, Ealet, Ebelke,
  Edmondson, Eisenstein, Ellsworth, Elsworth, Epstein, Eracleous, Escoffier,
  Esposito, Evans, Fan, Fern{\'a}ndez-Alvar, Feuillet, Filiz~Ak, Finley,
  Finoguenov, Flaherty, Fleming, Font-Ribera, Foster, Frinchaboy,
  Galbraith-Frew, Garc{\'\i}a, Garc{\'\i}a-Hern{\'a}ndez,
  Garc{\'\i}a~P{\'e}rez, Gaulme, Ge, G{\'e}nova-Santos, Georgakakis, Ghezzi,
  Gillespie, Girardi, Goddard, Gontcho, Gonz{\'a}lez~Hern{\'a}ndez, Grebel,
  Green, Niklas~Grieb, Grieves, Gunn, Guo, Harding, Hasselquist, Hawley,
  Hayden, Hearty, Hekker, Ho, Hogg, Holley-Bockelmann, Holtzman, Honscheid,
  Huber, Huehnerhoff, Ivans, Jiang, Johnson, Kinemuchi, Kirkby, Kitaura,
  Klaene, Knapp, Kneib, Koenig, Lam, Lan, Lang, Laurent, Le~Goff, Leauthaud,
  Lee, Lee, Licquia, Liu, Long, L{\'o}pez-Corredoira, Lorenzo-Oliveira,
  Lucatello, Lundgren, Lupton, Mack, Mahadevan, Maia, Majewski, Malanushenko,
  Malanushenko, Manchado, Manera, Mao, Maraston, Marchwinski, Margala, Martell,
  Martig, Masters, Mathur, McBride, McGehee, McGreer, McMahon, M{\'e}nard,
  Menzel, Merloni, M{\'e}sz{\'a}ros, Miller, Miralda-Escud{\'e}, Miyatake,
  Montero-Dorta, More, Morganson, Morice-Atkinson, Morrison, Mosser, Muna,
  Myers, Nandra, Newman, Neyrinck, Nguyen, Nichol, Nidever, Noterdaeme, Nuza,
  O{\textquoteright}Connell, O{\textquoteright}Connell,
  O{\textquoteright}Connell, Ogando, Olmstead, Oravetz, Oravetz, Osumi, Owen,
  Padgett, Padmanabhan, Paegert, Palanque-Delabrouille, Pan, Parejko,
  P{\^a}ris, Park, Pattarakijwanich, Pellejero-Ibanez, Pepper, Percival,
  P{\'e}rez-Fournon, Pe~rez Ra~fols, Petitjean, Pieri, Pinsonneault, Porto~de
  Mello, Prada, Prakash, Price-Whelan, Protopapas, Raddick, Rahman, Reid, Rich,
  Rix, Robin, Rockosi, Rodrigues, Rodr{\'\i}guez-Torres, Roe, Ross, Ross,
  Rossi, Ruan, Rubi{\~n}o-Mart{\'\i}n, Rykoff, Salazar-Albornoz, Salvato,
  Samushia, S{\'a}nchez, Santiago, Sayres, Schiavon, Schlegel, Schmidt,
  Schneider, Schultheis, Schwope, Sc{\'o}ccola, Scott, Sellgren, Seo,
  Serenelli, Shane, Shen, Shetrone, Shu, Silva~Aguirre, Sivarani, Skrutskie,
  Slosar, Smith, Sobreira, Souto, Stassun, Steinmetz, Stello, Strauss,
  Streblyanska, Suzuki, Swanson, Tan, Tayar, Terrien, Thakar, Thomas, Thomas,
  Thompson, Tinker, Tojeiro, Troup, Vargas-Maga{\~n}a, Vazquez, Verde, Viel,
  Vogt, Wake, Wang, Weaver, Weinberg, Weiner, White, Wilson, Wisniewski,
  Wood-Vasey, Ye~che, York, Zakamska, Zamora, Zasowski, Zehavi, Zhao, Zheng,
  Zhou, Zhou, Zou, \& Zhu}]{2015ApJS..219...12A}
Alam, S., Albareti, F.~D., Allende~Prieto, C., {et~al.} 2015, The Astrophysical
  Journal Supplement Series, 219, 12

\bibitem[{Allen {et~al.}(2007)Allen, Cruz, Koerner, McElwain, \&
  Reid}]{2007AJ....133..971A}
Allen, P.~R., Cruz, K.~K., Koerner, D.~W., McElwain, M.~W., \& Reid, N.~I.
  2007, The Astronomical Journal, 133, 971

\bibitem[{Allers \& Liu(2013)}]{2013ApJ...772...79A}
Allers, K.~N., \& Liu, M.~C. 2013, The Astrophysical Journal, 772, 79

\bibitem[{Alonso-Floriano {et~al.}(2015)Alonso-Floriano, Morales, Caballero,
  Montes, Klutsch, Mundt, Cort{\'e}s-Contreras, Ribas, Reiners, Amado,
  Quirrenbach, \& Jeffers}]{2015AA...577A.128A}
Alonso-Floriano, F.~J., Morales, J.~C., Caballero, J.~A., {et~al.} 2015,
  Astronomy and Astrophysics, 577, A128

\bibitem[{Babusiaux {et~al.}(2018)Babusiaux, van Leeuwen, Barstow, Jordi,
  Vallenari, Bossini, Bressan, {Gaia Collaboration}, \& {DPAC
  co-authors}}]{Babusiaux:2018di}
Babusiaux, C., van Leeuwen, F., Barstow, M., {et~al.} 2018, Astronomy {\&}
  Astrophysics, doi:10.1051/0004-6361/201832843

\bibitem[{Bardalez~Gagliuffi {et~al.}(2014)Bardalez~Gagliuffi, Burgasser,
  Gelino, Looper, Nicholls, Schmidt, Cruz, West, Gizis, \&
  Metchev}]{2014ApJ...794..143B}
Bardalez~Gagliuffi, D.~C., Burgasser, A.~J., Gelino, C.~R., {et~al.} 2014, The
  Astrophysical Journal, 794, 143

\bibitem[{Bell {et~al.}(2015)Bell, Mamajek, \& Naylor}]{2015MNRAS.454..593B}
Bell, C. P.~M., Mamajek, E.~E., \& Naylor, T. 2015, Monthly Notices of the
  Royal Astronomical Society, 454, 593

\bibitem[{B{\"o}hringer {et~al.}(2004)B{\"o}hringer, Schuecker, Guzzo, Collins,
  Voges, Cruddace, Ortiz-Gil, Chincarini, De~Grandi, Edge, MacGillivray,
  Neumann, Schindler, \& Shaver}]{2004AA...425..367B}
B{\"o}hringer, H., Schuecker, P., Guzzo, L., {et~al.} 2004, Astronomy and
  Astrophysics, 425, 367

\bibitem[{Boller {et~al.}(2016)Boller, Freyberg, Tr{\"u}mper, Haberl, Voges, \&
  Nandra}]{2016AA...588A.103B}
Boller, T., Freyberg, M.~J., Tr{\"u}mper, J., {et~al.} 2016, Astronomy {\&}
  Astrophysics, 588, A103

\bibitem[{Bouvier {et~al.}(2008)Bouvier, Kendall, Meeus, Testi, Moraux,
  Stauffer, James, Cuillandre, Irwin, McCaughrean, Baraffe, \&
  Bertin}]{2008AA...481..661B}
Bouvier, J., Kendall, T.~R., Meeus, G., {et~al.} 2008, Astronomy and
  Astrophysics, 481, 661

\bibitem[{Bowler {et~al.}(2015)Bowler, Liu, Shkolnik, \&
  Tamura}]{2015ApJS..216....7B}
Bowler, B.~P., Liu, M.~C., Shkolnik, E.~L., \& Tamura, M. 2015, The
  Astrophysical Journal Supplement, 216, 7

\bibitem[{Bowler {et~al.}(2017)Bowler, Liu, Mawet, Ngo, Malo, Mace, McLane, Lu,
  Tristan, Hinkley, Hillenbrand, Shkolnik, Benneke, \&
  Best}]{2017AJ....153...18B}
Bowler, B.~P., Liu, M.~C., Mawet, D., {et~al.} 2017, The Astronomical Journal,
  153, 18

\bibitem[{Brandt \& Huang(2015)}]{2015ApJ...807...24B}
Brandt, T.~D., \& Huang, C.~X. 2015, The Astrophysical Journal, 807, 24

\bibitem[{Chambers {et~al.}(2016)Chambers, Magnier, Metcalfe, Flewelling,
  Huber, Waters, Denneau, Draper, Farrow, Finkbeiner, Holmberg, Koppenhoefer,
  Price, Saglia, Schlafly, Smartt, Sweeney, Wainscoat, Burgett, Grav, Heasley,
  Hodapp, Jedicke, Kaiser, Kudritzki, Luppino, Lupton, Monet, Morgan, Onaka,
  Stubbs, Tonry, Banados, Bell, Bender, Bernard, Botticella, Casertano,
  Chastel, Chen, Chen, Cole, Deacon, Frenk, Fitzsimmons, Gezari, Goessl,
  Goggia, Goldman, Grebel, Hambly, Hasinger, Heavens, Heckman, Henderson,
  Henning, Holman, Hopp, Ip, Isani, Keyes, Koekemoer, Kotak, Long, Lucey, Liu,
  Martin, McLean, Morganson, Murphy, Nieto-Santisteban, Norberg, Peacock, Pier,
  Postman, Primak, Rae, Rest, Riess, Riffeser, Rix, R{\"o}ser, Schilbach,
  Schultz, Scolnic, Szalay, Seitz, Shiao, Small, Smith, Soderblom, Taylor,
  Thakar, Thiel, Thilker, Urata, Valenti, Walter, Watters, Werner, White,
  Wood-Vasey, \& Wyse}]{2016arXiv161205560C}
Chambers, K.~C., Magnier, E.~A., Metcalfe, N., {et~al.} 2016, arXiv.org,
  arXiv:1612.05560

\bibitem[{Choi {et~al.}(2016)Choi, Dotter, Conroy, Cantiello, Paxton, \&
  Johnson}]{2016ApJ...823..102C}
Choi, J., Dotter, A., Conroy, C., {et~al.} 2016, The Astrophysical Journal,
  823, 102

\bibitem[{Comer{\'o}n {et~al.}(2013)Comer{\'o}n, Spezzi, Lopez~Mart{\'\i}, \&
  Mer{\'\i}n}]{2013AA...554A..86C}
Comer{\'o}n, F., Spezzi, L., Lopez~Mart{\'\i}, B., \& Mer{\'\i}n, B. 2013,
  Astronomy and Astrophysics, 554, A86

\bibitem[{Cook {et~al.}(2016)Cook, Pinfield, Marocco, Burningham, Jones, Frith,
  Zhong, Luo, Qi, Lucas, Gromadzki, Day-Jones, Kurtev, Guo, Wang, Bai, Yi, \&
  Smart}]{2016MNRAS.457.2192C}
Cook, N.~J., Pinfield, D.~J., Marocco, F., {et~al.} 2016, Monthly Notices of
  the Royal Astronomical Society, 457, 2192

\bibitem[{Cropper {et~al.}(2018)Cropper, Katz, Sartoretti, \& {et
  al.}}]{Cropper:2018jx}
Cropper, M., Katz, D., Sartoretti, P., \& {et al.} 2018, Astronomy {\&}
  Astrophysics, doi:10.1051/0004-6361/201832763

\bibitem[{Cruz {et~al.}(2003)Cruz, Lowrance, Reid, Liebert, \&
  Kirkpatrick}]{2003AJ....126.2421C}
Cruz, K.~K., Lowrance, P., Reid, N.~I., Liebert, J., \& Kirkpatrick, D.~J.
  2003, The Astronomical Journal, 126, 2421

\bibitem[{Cruz \& Reid(2002)}]{2002AJ....123.2828C}
Cruz, K.~K., \& Reid, N.~I. 2002, The Astronomical Journal, 123, 2828

\bibitem[{Cruz {et~al.}(2007)Cruz, Reid, Kirkpatrick, Burgasser, Liebert,
  Solomon, Schmidt, Allen, Hawley, \& Covey}]{2007AJ....133..439C}
Cruz, K.~L., Reid, I.~N., Kirkpatrick, J.~D., {et~al.} 2007, The Astronomical
  Journal, 133, 439

\bibitem[{Davenport {et~al.}(2012)Davenport, Becker, Kowalski, Hawley, Schmidt,
  Hilton, Sesar, \& Cutri}]{2012ApJ...748...58D}
Davenport, J. R.~A., Becker, A.~C., Kowalski, A.~F., {et~al.} 2012, The
  Astrophysical Journal, 748, 58

\bibitem[{Deshpande {et~al.}(2013)Deshpande, Blake, Bender, Mahadevan, Terrien,
  Carlberg, Zasowski, Crepp, Rajpurohit, Reyl{\'e}, Nidever, Schneider,
  Allende~Prieto, Bizyaev, Ebelke, Fleming, Frinchaboy, Ge, Hearty,
  Hern{\'a}ndez, Malanushenko, Malanushenko, Majewski, Marchwinski, Muna,
  Oravetz, Pan, Schiavon, Shetrone, Simmons, Stassun, Wilson, \&
  Wisniewski}]{2013AJ....146..156D}
Deshpande, R., Blake, C.~H., Bender, C.~F., {et~al.} 2013, The Astronomical
  Journal, 146, 156

\bibitem[{Desidera {et~al.}(2015)Desidera, Covino, Messina, Carson, Hagelberg,
  Schlieder, Biazzo, Alcal{\'a}, Chauvin, Vigan, Beuzit, Bonavita, Bonnefoy,
  Delorme, D'Orazi, Esposito, Feldt, Girardi, Gratton, Henning, Lagrange,
  Lanzafame, Launhardt, Marmier, Melo, Meyer, Mouillet, Moutou, S{\'e}gransan,
  Udry, \& Zaidi}]{2015AA...573A.126D}
Desidera, S., Covino, E., Messina, S., {et~al.} 2015, Astronomy {\&}
  Astrophysics, 573, A126

\bibitem[{Desrochers {et~al.}(2018)Desrochers, Artigau, Gagn{\'e}, Doyon, Malo,
  Faherty, \& Lafreni{\`e}re}]{2018ApJ...852...55D}
Desrochers, M.-E., Artigau, {\'E}., Gagn{\'e}, J., {et~al.} 2018, The
  Astrophysical Journal, 852, 55

\bibitem[{Dieterich {et~al.}(2014)Dieterich, Henry, Jao, Winters, Hosey,
  Riedel, \& Subasavage}]{2014AJ....147...94D}
Dieterich, S.~B., Henry, T.~J., Jao, W.-C., {et~al.} 2014, The Astronomical
  Journal, 147, 94

\bibitem[{Duflot {et~al.}(1995)Duflot, Figon, \&
  Meyssonnier}]{1995AAS..114..269D}
Duflot, M., Figon, P., \& Meyssonnier, N. 1995, Astronomy and Astrophysics
  Supplement, 114, 269

\bibitem[{Evans(1967)}]{1967IAUS...30...57E}
Evans, D.~S. 1967, Determination of Radial Velocities and their Applications,
  30, 57

\bibitem[{Evans {et~al.}(1964)Evans, Laing, Menzies, \&
  Stoy}]{1964RGOB...85..207E}
Evans, D.~S., Laing, J.~D., Menzies, A., \& Stoy, R.~H. 1964, Royal Observatory
  bulletins ; no. 85; Royal Observatory bulletins. Series E ; no. 85., 85, 207

\bibitem[{Evans {et~al.}(2018)Evans, Riello, De~Angeli, Carrasco, Montegriffo,
  Fabricius, Jordi, Palaversa, Diener, Busso, Weiler, Cacciari, \& van
  Leeuwen}]{Evans:2018cj}
Evans, D.~W., Riello, M., De~Angeli, F., {et~al.} 2018, Astronomy {\&}
  Astrophysics, doi:10.1051/0004-6361/201832756

\bibitem[{Faherty {et~al.}(2009)Faherty, Cruz, Burgasser, Shara, Walter, \&
  Gelino}]{2009AJ....137....1F}
Faherty, J.~K., Cruz, K.~K., Burgasser, A.~J., {et~al.} 2009, The Astronomical
  Journal, 137, 1

\bibitem[{Faherty {et~al.}(2018)Faherty, Gagn{\'e}, Burgasser, Mamajek,
  Gonzales, Bardalez-Gagliuffi, \& Marocco}]{2018arXiv180501573F}
Faherty, J.~K., Gagn{\'e}, J., Burgasser, A.~J., {et~al.} 2018, arXiv.org,
  arXiv:1805.01573

\bibitem[{Faherty {et~al.}(2016)Faherty, Riedel, Cruz, Gagn{\'e}, Filippazzo,
  Lambrides, Fica, Weinberger, Thorstensen, Tinney, Baldassare, Lemonier, \&
  Rice}]{2016ApJS..225...10F}
Faherty, J.~K., Riedel, A.~R., Cruz, K.~K., {et~al.} 2016, The Astrophysical
  Journal Supplement Series, 225, 10

\bibitem[{Fleming {et~al.}(1988)Fleming, Liebert, Gioia, \&
  Maccacaro}]{1988ApJ...331..958F}
Fleming, T.~A., Liebert, J., Gioia, I.~M., \& Maccacaro, T. 1988, The
  Astrophysical Journal, 331, 958

\bibitem[{Gagn{\'e} {et~al.}(2018{\natexlab{a}})Gagn{\'e}, Faherty, \&
  Fontaine}]{2018arXiv180405248G}
Gagn{\'e}, J., Faherty, J.~K., \& Fontaine, G. 2018{\natexlab{a}}, arXiv.org,
  arXiv:1804.05248

\bibitem[{Gagn{\'e} {et~al.}(2015{\natexlab{a}})Gagn{\'e}, Lafreni{\`e}re,
  Doyon, Malo, \& Artigau}]{2015ApJ...798...73G}
Gagn{\'e}, J., Lafreni{\`e}re, D., Doyon, R., Malo, L., \& Artigau, {\'E}.
  2015{\natexlab{a}}, The Astrophysical Journal, 798, 73

\bibitem[{Gagn{\'e} {et~al.}(2018{\natexlab{b}})Gagn{\'e}, Roy-Loubier,
  Faherty, Doyon, \& Malo}]{Gagne:2018un}
Gagn{\'e}, J., Roy-Loubier, O., Faherty, J.~K., Doyon, R., \& Malo, L.
  2018{\natexlab{b}}, arXiv.org, 1804.03093v1

\bibitem[{Gagn\'e {et~al.}(2018)Gagn\'e, Schneider, \&
  Cushing}]{zenodofindercharts}
Gagn\'e, J., Schneider, A., \& Cushing, M. 2018, {Finder Charts Python Package
  v1.0, Zenodo}, doi:10.5281/zenodo.1237017

\bibitem[{Gagn{\'e} {et~al.}(2015{\natexlab{b}})Gagn{\'e}, Faherty, Cruz,
  Lafreni{\`e}re, Doyon, Malo, Burgasser, Naud, Artigau, Bouchard, Gizis, \&
  Albert}]{2015ApJS..219...33G}
Gagn{\'e}, J., Faherty, J.~K., Cruz, K.~L., {et~al.} 2015{\natexlab{b}}, The
  Astrophysical Journal Supplement Series, 219, 33

\bibitem[{Gagn{\'e} {et~al.}(2017)Gagn{\'e}, Faherty, Mamajek, Malo, Doyon,
  Filippazzo, Weinberger, Donaldson, L{\'e}pine, Lafreni{\`e}re, Artigau,
  Burgasser, Looper, Boucher, Beletsky, Camnasio, Brunette, \&
  Arboit}]{2017ApJS..228...18G}
Gagn{\'e}, J., Faherty, J.~K., Mamajek, E.~E., {et~al.} 2017, The Astrophysical
  Journal Supplement Series, 228, 18

\bibitem[{Gagn{\'e} {et~al.}(2018)Gagn{\'e}, Mamajek, Malo, Riedel, Rodriguez,
  Lafreni{\`e}re, Faherty, Roy-Loubier, Pueyo, Robin, \&
  Doyon}]{2018ApJ...856...23G}
Gagn{\'e}, J., Mamajek, E.~E., Malo, L., {et~al.} 2018, The Astrophysical
  Journal, 856, 23

\bibitem[{{Gaia Collaboration} {et~al.}(2018){Gaia Collaboration}, Brown,
  Vallenari, Prusti, de~Bruijne, \& {et al.}}]{GaiaCollaboration:2018io}
{Gaia Collaboration}, Brown, A. G.~A., Vallenari, A., {et~al.} 2018, Astronomy
  {\&} Astrophysics, doi:10.1051/0004-6361/201833051

\bibitem[{{Gaia Collaboration} {et~al.}(2016){Gaia Collaboration}, Brown,
  Vallenari, Prusti, de~Bruijne, Mignard, Drimmel, Babusiaux, Bailer-Jones,
  Bastian, Biermann, Evans, Eyer, Jansen, Jordi, Katz, Klioner, Lammers,
  Lindegren, Luri, O'Mullane, Panem, Pourbaix, Randich, Sartoretti, Siddiqui,
  Soubiran, Valette, van Leeuwen, Walton, Aerts, Arenou, Cropper, H{\o}g,
  Lattanzi, Grebel, Holland, Huc, Passot, Perryman, Bramante, Cacciari,
  Casta{\~n}eda, Chaoul, Cheek, De~Angeli, Fabricius, Guerra, Hern{\'a}ndez,
  Jean-Antoine-Piccolo, Masana, Messineo, Mowlavi, Nienartowicz,
  Ord{\'o}{\~n}ez-Blanco, Panuzzo, Portell, Richards, Riello, Seabroke, Tanga,
  Th{\'e}venin, Torra, Els, Gracia-Abril, Comoretto, Garcia-Reinaldos, Lock,
  Mercier, Altmann, Andrae, Astraatmadja, Bellas-Velidis, Benson, Berthier,
  Blomme, Busso, Carry, Cellino, Clementini, Cowell, Creevey, Cuypers,
  Davidson, De~Ridder, de~Torres, Delchambre, Dell'Oro, Ducourant, Fr{\'e}mat,
  Garc{\'\i}a-Torres, Gosset, Halbwachs, Hambly, Harrison, Hauser, Hestroffer,
  Hodgkin, Huckle, Hutton, Jasniewicz, Jordan, Kontizas, Korn, Lanzafame,
  Manteiga, Moitinho, Muinonen, Osinde, Pancino, Pauwels, Petit, Recio-Blanco,
  Robin, Sarro, Siopis, Smith, Smith, Sozzetti, Thuillot, van Reeven, Viala,
  Abbas, Abreu~Aramburu, Accart, Aguado, Allan, Allasia, Altavilla,
  {\'A}lvarez, Alves, Anderson, Andrei, Anglada~Varela, Antiche, Antoja, Anton,
  Arcay, Bach, Baker, Balaguer-N{\'u}{\~n}ez, Barache, Barata, Barbier,
  Barblan, Barrado~y Navascu{\'e}s, Barros, Barstow, Becciani, Bellazzini,
  Bello~Garc{\'\i}a, Belokurov, Bendjoya, Berihuete, Bianchi, Bienaym{\'e},
  Billebaud, Blagorodnova, Blanco-Cuaresma, Boch, Bombrun, Borrachero,
  Bouquillon, Bourda, Bouy, Bragaglia, Breddels, Brouillet, Br{\"u}semeister,
  Bucciarelli, Burgess, Burgon, Burlacu, Busonero, Buzzi, Caffau, Cambras,
  Campbell, Cancelliere, Cantat-Gaudin, Carlucci, Carrasco, Castellani,
  Charlot, Charnas, Chiavassa, Clotet, Cocozza, Collins, Costigan, Crifo,
  Cross, Crosta, Crowley, Dafonte, Damerdji, Dapergolas, David, David, De~Cat,
  de~Felice, de~Laverny, De~Luise, De~March, de~Martino, de~Souza, Debosscher,
  del Pozo, Delbo, Delgado, Delgado, Di~Matteo, Diakite, Distefano, Dolding,
  Dos~Anjos, Drazinos, Dur{\'a}n, Dzigan, Edvardsson, Enke, Evans,
  Eynard~Bontemps, Fabre, Fabrizio, Faigler, Falc{\~a}o, Farr{\`a}s~Casas,
  Federici, Fedorets, Fern{\'a}ndez-Hern{\'a}ndez, Fernique, Fienga, Figueras,
  Filippi, Findeisen, Fonti, Fouesneau, Fraile, Fraser, Fuchs, Gai, Galleti,
  Galluccio, Garabato, Garc{\'\i}a-Sedano, Garofalo, Garralda, Gavras, Gerssen,
  Geyer, Gilmore, Girona, Giuffrida, Gomes, Gonz{\'a}lez-Marcos,
  Gonz{\'a}lez-N{\'u}{\~n}ez, Gonz{\'a}lez-Vidal, Granvik, Guerrier, Guillout,
  Guiraud, G{\'u}rpide, Guti{\'e}rrez-S{\'a}nchez, {Guy, L. P.}, Haigron,
  Hatzidimitriou, Haywood, Heiter, Helmi, Hobbs, Hofmann, Holl, Holland, Hunt,
  Hypki, Icardi, Irwin, Jevardat~de Fombelle, Jofr{\'e}, Jonker, Jorissen,
  Julbe, Karampelas, Kochoska, Kohley, Kolenberg, Kontizas, Koposov,
  Kordopatis, Koubsky, Krone-Martins, Kudryashova, Kull, Bachchan,
  Lacoste-Seris, Lanza, Lavigne, Le~Poncin-Lafitte, Lebreton, Lebzelter,
  Leccia, Leclerc, Lecoeur-Taibi, Lemaitre, Lenhardt, Leroux, Liao, Licata,
  Lindstr{\o}m, Lister, Livanou, Lobel, L{\"o}ffler, L{\'o}pez, Lorenz,
  MacDonald, Magalh{\~a}es~Fernandes, Managau, Mann, Mantelet, Marchal,
  Marchant, Marconi, Marinoni, Marrese, Marschalk{\'o}, Marshall,
  Mart{\'\i}n-Fleitas, Martino, Mary, Matijevi{\v c}, Mazeh, McMillan, Messina,
  Michalik, Millar, Miranda, Molina, Molinaro, Molinaro, Moln{\'a}r, Moniez,
  Montegriffo, Mor, Mora, Morbidelli, Morel, Morgenthaler, Morris, Mulone,
  Muraveva, Musella, Narbonne, Nelemans, Nicastro, Noval, Ord{\'e}novic,
  Ordieres-Mer{\'e}, Osborne, Pagani, Pagano, Pailler, Palacin, Palaversa,
  Parsons, Pecoraro, Pedrosa, Pentik{\"a}inen, Pichon, Piersimoni, Pineau,
  Plachy, Plum, Poujoulet, Pr{\v s}a, Pulone, Ragaini, Rago, Rambaux,
  Ramos-Lerate, Ranalli, Rauw, Read, Regibo, Reyl{\'e}, Ribeiro, Rimoldini,
  Ripepi, Riva, Rixon, Roelens, Romero-G{\'o}mez, Rowell, Royer, Ruiz-Dern,
  Sadowski, Sagrist{\`a}~Sell{\'e}s, Sahlmann, Salgado, Salguero, Sarasso,
  Savietto, Schultheis, Sciacca, Segol, Segovia, S{\'e}gransan, Shih,
  Smareglia, Smart, Solano, Solitro, Sordo, Soria~Nieto, Souchay, Spagna,
  Spoto, Stampa, Steele, Steidelm{\"u}ller, Stephenson, Stoev, Suess,
  S{\"u}veges, Surdej, Szabados, Szegedi-Elek, Tapiador, Taris, Tauran, Taylor,
  Teixeira, Terrett, Tingley, Trager, Turon, Ulla, Utrilla, Valentini, van
  Elteren, Van~Hemelryck, van Leeuwen, Varadi, Vecchiato, Veljanoski, Via,
  Vicente, Vogt, Voss, Votruba, Voutsinas, Walmsley, Weiler, Weingrill, Wevers,
  Wyrzykowski, Yoldas, {\v Z}erjal, Zucker, Zurbach, Zwitter, Alecu, Allen,
  Allende~Prieto, Amorim, Anglada-Escude, Arsenijevic, Azaz, Balm, Beck,
  Bernstein, Bigot, Bijaoui, Blasco, Bonfigli, Bono, Boudreault, Bressan,
  Brown, Brunet, Bunclark, Buonanno, Butkevich, Carret, Carrion, Chemin,
  Ch{\'e}reau, Corcione, Darmigny, de~Boer, de~Teodoro, de~Zeeuw, Delle~Luche,
  Domingues, Dubath, Fodor, Fr{\'e}zouls, Fries, Fustes, Fyfe, Gallardo,
  Gallegos, Gardiol, Gebran, Gomboc, Gomez, Grux, Gueguen, Heyrovsky, Hoar,
  Iannicola, Isasi~Parache, Janotto, Joliet, Jonckheere, Keil, Kim, Klagyivik,
  Klar, Knude, Kochukhov, Kolka, Kos, Kutka, Lainey, LeBouquin, Liu, Loreggia,
  Makarov, Marseille, Martayan, Martinez-Rubi, Massart, Meynadier, Mignot,
  Munari, Nguyen, Nordlander, Ocvirk, O'Flaherty, Olias~Sanz, Ortiz, Osorio,
  Oszkiewicz, Ouzounis, Palmer, Park, Pasquato, Peltzer, Peralta, P{\'e}turaud,
  Pieniluoma, Pigozzi, Poels, Prat, Prod'homme, Raison, Rebordao, Risquez,
  Rocca-Volmerange, Rosen, Ruiz-Fuertes, Russo, Sembay, Serraller~Vizcaino,
  Short, Siebert, Silva, Sinachopoulos, Slezak, Soffel, Sosnowska, Strai{\v
  z}ys, ter Linden, Terrell, Theil, Tiede, Troisi, Tsalmantza, Tur, Vaccari,
  Vachier, Valles, Van~Hamme, Veltz, Virtanen, Wallut, Wichmann, Wilkinson,
  Ziaeepour, \& Zschocke}]{2016AA...595A...2G}
---. 2016, Astronomy and Astrophysics, 595, A2

\bibitem[{Gaidos {et~al.}(2014)Gaidos, Mann, L{\'e}pine, Buccino, James,
  Ansdell, Petrucci, Mauas, \& Hilton}]{2014MNRAS.443.2561G}
Gaidos, E., Mann, A.~W., L{\'e}pine, S., {et~al.} 2014, Monthly Notices of the
  Royal Astronomical Society, 443, 2561

\bibitem[{Galicher {et~al.}(2016)Galicher, Marois, Macintosh, Zuckerman,
  Barman, Konopacky, Song, Patience, Lafreniere, Doyon, \&
  Nielsen}]{2016AA...594A..63G}
Galicher, R., Marois, C., Macintosh, B., {et~al.} 2016, Astronomy and
  Astrophysics, 594, A63

\bibitem[{Gontcharov(2006)}]{2006AstL...32..759G}
Gontcharov, G.~A. 2006, Astronomy Letters, 32, 759

\bibitem[{Gray {et~al.}(2006)Gray, Corbally, Garrison, McFadden, Bubar,
  McGahee, O'Donoghue, \& Knox}]{2006AJ....132..161G}
Gray, R.~O., Corbally, C.~J., Garrison, R.~F., {et~al.} 2006, The Astronomical
  Journal, 132, 161

\bibitem[{Gray {et~al.}(2003)Gray, Corbally, Garrison, McFadden, \&
  Robinson}]{2003AJ....126.2048G}
Gray, R.~O., Corbally, C.~J., Garrison, R.~F., McFadden, M.~T., \& Robinson,
  P.~E. 2003, The Astronomical Journal, 126, 2048

\bibitem[{Guieu {et~al.}(2006)Guieu, Dougados, Monin, Magnier, \&
  Mart{\'\i}n}]{2006AA...446..485G}
Guieu, S., Dougados, C., Monin, J.-L., Magnier, E., \& Mart{\'\i}n, E.~L. 2006,
  Astronomy and Astrophysics, 446, 485

\bibitem[{Hambly {et~al.}(2018)Hambly, Cropper, Boudreault, Crowley, Kohley,
  de~Bruijne, Dolding, Fabricius, \& {et al.}}]{Hambly:2018gr}
Hambly, N.~C., Cropper, M., Boudreault, S., {et~al.} 2018, Astronomy {\&}
  Astrophysics, doi:10.1051/0004-6361/201832716

\bibitem[{Hawley {et~al.}(2002)Hawley, Covey, Knapp, Golimowski, Fan, Anderson,
  Gunn, Harris, Ivezi{\'c}, Long, Lupton, McGehee, Narayanan, Peng, Schlegel,
  Schneider, Spahn, Strauss, Szkody, Tsvetanov, Walkowicz, Brinkmann, Harvanek,
  Hennessy, Kleinman, Krzesinski, Long, Neilsen, Newman, Nitta, Snedden, \&
  York}]{2002AJ....123.3409H}
Hawley, S.~L., Covey, K.~R., Knapp, G.~R., {et~al.} 2002, The Astronomical
  Journal, 123, 3409

\bibitem[{Herbig {et~al.}(1986)Herbig, Vrba, \& Rydgren}]{1986AJ.....91..575H}
Herbig, G.~H., Vrba, F.~J., \& Rydgren, A.~E. 1986, Astronomical Journal, 91,
  575

\bibitem[{Houk(1982)}]{1982mcts.book.....H}
Houk, N. 1982, Michigan Catalogue of Two-dimensional Spectral Types for the HD
  stars. Volume 3. Declinations -40 to -26.

\bibitem[{Houk \& Cowley(1975)}]{1975mcts.book.....H}
Houk, N., \& Cowley, A.~P. 1975, University of Michigan Catalogue of
  two-dimensional spectral types for the HD stars. Volume I. Declinations -90
  to -53., I

\bibitem[{Houk \& Smith-Moore(1988)}]{1988mcts.book.....H}
Houk, N., \& Smith-Moore, M. 1988, Michigan Catalogue of Two-dimensional
  Spectral Types for the HD Stars. Volume 4, 4

\bibitem[{Jeffries(1995)}]{1995MNRAS.273..559J}
Jeffries, R.~D. 1995, Monthly Notices of the Royal Astronomical Society, 273,
  559

\bibitem[{Jones \& West(2016)}]{2016ApJ...817....1J}
Jones, D.~O., \& West, A.~A. 2016, The Astrophysical Journal, 817, 1

\bibitem[{Jones {et~al.}(2015)Jones, White, Boyajian, Schaefer, Baines,
  Ireland, Patience, McAlister, \& Ten~Brummelaar}]{2015AAS...22511203J}
Jones, J., White, R.~J., Boyajian, T.~S., {et~al.} 2015, American Astronomical
  Society, 225, 112.03

\bibitem[{Kenyon \& Hartmann(1995)}]{1995ApJS..101..117K}
Kenyon, S.~J., \& Hartmann, L. 1995, Astrophysical Journal Supplement v.101,
  101, 117

\bibitem[{Kirkpatrick {et~al.}(1991)Kirkpatrick, Henry, \&
  McCarthy}]{1991ApJS...77..417K}
Kirkpatrick, D.~J., Henry, T.~J., \& McCarthy, D. W.~J. 1991, Astrophysical
  Journal Supplement Series, 77, 417

\bibitem[{Kirkpatrick {et~al.}(2008)Kirkpatrick, Lowrance, Cruz, Barman,
  Burgasser, Looper, Tinney, Gelino, Liebert, Carpenter, Hillenbrand, \&
  Stauffer}]{2008ApJ...689.1295K}
Kirkpatrick, D.~J., Lowrance, P., Cruz, K.~K., {et~al.} 2008, The Astrophysical
  Journal, 689, 1295

\bibitem[{Koen {et~al.}(2010)Koen, Kilkenny, van Wyk, \&
  Marang}]{2010MNRAS.403.1949K}
Koen, C., Kilkenny, D., van Wyk, F., \& Marang, F. 2010, Monthly Notices of the
  Royal Astronomical Society, 403, 1949

\bibitem[{Kraus {et~al.}(2014)Kraus, Shkolnik, Allers, \&
  Liu}]{2014AJ....147..146K}
Kraus, A.~L., Shkolnik, E.~L., Allers, K.~N., \& Liu, M.~C. 2014, The
  Astronomical Journal, 147, 146

\bibitem[{Krautter {et~al.}(1997)Krautter, Wichmann, Schmitt, Alcal{\'a},
  Neuh{\"a}user, \& Terranegra}]{1997AAS..123..329K}
Krautter, J., Wichmann, R., Schmitt, J. H. M.~M., {et~al.} 1997, A {\&} A
  Supplement series, 123, 329

\bibitem[{Kunder {et~al.}(2017)Kunder, Kordopatis, Steinmetz, Zwitter,
  McMillan, Casagrande, Enke, Wojno, Valentini, Chiappini, Matijevi{\v c},
  Siviero, de~Laverny, Recio-Blanco, Bijaoui, Wyse, Binney, Grebel, Helmi,
  Jofre, Antoja, Gilmore, Siebert, Famaey, Bienaym{\'e}, Gibson, Freeman,
  Navarro, Munari, Seabroke, Anguiano, {\v Z}erjal, Minchev, Reid,
  Bland-Hawthorn, Kos, Sharma, Watson, Parker, Scholz, Burton, Cass, Hartley,
  Fiegert, Stupar, Ritter, Hawkins, Gerhard, Chaplin, Davies, Elsworth, Lund,
  Miglio, \& Mosser}]{2017AJ....153...75K}
Kunder, A., Kordopatis, G., Steinmetz, M., {et~al.} 2017, The Astronomical
  Journal, 153, 75

\bibitem[{Law {et~al.}(2008)Law, Hodgkin, \& Mackay}]{2008MNRAS.384..150L}
Law, N.~M., Hodgkin, S.~T., \& Mackay, C.~D. 2008, Monthly Notices of the Royal
  Astronomical Society, 384, 150

\bibitem[{Lawrence {et~al.}(2007)Lawrence, Warren, Almaini, Edge, Hambly,
  Jameson, Lucas, Casali, Adamson, Dye, Emerson, Foucaud, Hewett, Hirst,
  Hodgkin, Irwin, Lodieu, McMahon, Simpson, Smail, Mortlock, \&
  Folger}]{2007MNRAS.379.1599L}
Lawrence, A., Warren, S.~J., Almaini, O., {et~al.} 2007, Monthly Notices of the
  Royal Astronomical Society, 379, 1599

\bibitem[{L{\'e}pine {et~al.}(2013)L{\'e}pine, Hilton, Mann, Wilde,
  Rojas-Ayala, Cruz, \& Gaidos}]{2013AJ....145..102L}
L{\'e}pine, S., Hilton, E.~J., Mann, A.~W., {et~al.} 2013, The Astronomical
  Journal, 145, 102

\bibitem[{Lindegren {et~al.}(2018)Lindegren, Hern{\'a}ndez, Bombrun, Klioner,
  Bastian, \& Ramos-Lerate}]{Lindegren:2018gy}
Lindegren, L., Hern{\'a}ndez, J., Bombrun, A., {et~al.} 2018, Astronomy {\&}
  Astrophysics, doi:10.1051/0004-6361/201832727

\bibitem[{Luhman(2006)}]{2006ApJ...645..676L}
Luhman, K.~L. 2006, The Astrophysical Journal, 645, 676

\bibitem[{Luhman {et~al.}(2005)Luhman, Stauffer, \&
  Mamajek}]{2005ApJ...628L..69L}
Luhman, K.~L., Stauffer, J.~R., \& Mamajek, E.~E. 2005, The Astrophysical
  Journal, 628, L69

\bibitem[{Luri {et~al.}(2018)Luri, A~Brown, Sarro, Arenou, Bailer-Jones,
  Castro-Ginard, de~Bruijne, Prusti, Babusiaux, \& Delgado}]{Luri:2018eu}
Luri, X., A~Brown, A.~G., Sarro, L., {et~al.} 2018, Astronomy {\&}
  Astrophysics, doi:10.1051/0004-6361/201832964

\bibitem[{Malo {et~al.}(2014{\natexlab{a}})Malo, Artigau, Doyon,
  Lafreni{\`e}re, Albert, \& Gagn{\'e}}]{2014ApJ...788...81M}
Malo, L., Artigau, {\'E}., Doyon, R., {et~al.} 2014{\natexlab{a}}, The
  Astrophysical Journal, 788, 81

\bibitem[{Malo {et~al.}(2014{\natexlab{b}})Malo, Doyon, Feiden, Albert,
  Lafreni{\`e}re, Artigau, Gagn{\'e}, \& Riedel}]{2014ApJ...792...37M}
Malo, L., Doyon, R., Feiden, G.~A., {et~al.} 2014{\natexlab{b}}, The
  Astrophysical Journal, 792, 37

\bibitem[{Malo {et~al.}(2013)Malo, Doyon, Lafreni{\`e}re, Artigau, Gagn{\'e},
  Baron, \& Riedel}]{2013ApJ...762...88M}
Malo, L., Doyon, R., Lafreni{\`e}re, D., {et~al.} 2013, The Astrophysical
  Journal, 762, 88

\bibitem[{Mamajek(2016)}]{mamajek118tau}
Mamajek, E. 2016, {A New Candidate Young Stellar Group at d=121 pc Associated
  with 118 Tauri}, doi:10.6084/m9.figshare.3122689.v1

\bibitem[{Marois {et~al.}(2008)Marois, Macintosh, Barman, Zuckerman, Song,
  Patience, Lafreniere, \& Doyon}]{2008Sci...322.1348M}
Marois, C., Macintosh, B., Barman, T.~S., {et~al.} 2008, Science, 322, 1348

\bibitem[{Martin {et~al.}(2005)Martin, Fanson, Schiminovich, Morrissey,
  Friedman, Barlow, Conrow, Grange, Jelinsky, Milliard, Siegmund, Bianchi,
  Byun, Donas, Forster, Heckman, Lee, Madore, Malina, Neff, Rich, Small,
  Surber, Szalay, Welsh, \& Wyder}]{2005ApJ...619L...1M}
Martin, D.~C., Fanson, J., Schiminovich, D., {et~al.} 2005, The Astrophysical
  Journal, 619, L1

\bibitem[{McMahon {et~al.}(2013)McMahon, Banerji, Gonzalez, Koposov, Bejar,
  Lodieu, Rebolo, \& Collaboration}]{2013Msngr.154...35M}
McMahon, R.~G., Banerji, M., Gonzalez, E., {et~al.} 2013, The Messenger, 154,
  35

\bibitem[{Mignard {et~al.}(2018)Mignard, Klioner, Lindegren, Hern{\'a}ndez,
  Bastian, Bombrun, Hobbs, Lammers, \& {et al.}}]{Mignard:2018bj}
Mignard, F., Klioner, S., Lindegren, L., {et~al.} 2018, Astronomy {\&}
  Astrophysics, doi:10.1051/0004-6361/201832916

\bibitem[{Murphy \& Lawson(2015)}]{2015MNRAS.447.1267M}
Murphy, S.~J., \& Lawson, W.~A. 2015, Monthly Notices of the Royal Astronomical
  Society, 447, 1267

\bibitem[{Murphy {et~al.}(2013)Murphy, Lawson, \&
  Bessell}]{2013MNRAS.435.1325M}
Murphy, S.~J., Lawson, W.~A., \& Bessell, M.~S. 2013, Monthly Notices of the
  Royal Astronomical Society, 435, 1325

\bibitem[{Newton {et~al.}(2014)Newton, Charbonneau, Irwin, Berta-Thompson,
  Rojas-Ayala, Covey, \& Lloyd}]{2014AJ....147...20N}
Newton, E.~R., Charbonneau, D., Irwin, J., {et~al.} 2014, The Astronomical
  Journal, 147, 20

\bibitem[{Ochsenbein {et~al.}(2000)Ochsenbein, Bauer, \&
  Marcout}]{2000AAS..143...23O}
Ochsenbein, F., Bauer, P., \& Marcout, J. 2000, Astronomy and Astrophysics
  Supplement, 143, 23

\bibitem[{Pecaut \& Mamajek(2016)}]{2016MNRAS.461..794P}
Pecaut, M.~J., \& Mamajek, E.~E. 2016, Monthly Notices of the Royal
  Astronomical Society, 461, 794

\bibitem[{Perryman {et~al.}(1997)Perryman, Lindegren, Kovalevsky, Hoeg,
  Bastian, Bernacca, Cr{\'e}z{\'e}, Donati, Grenon, Grewing, van Leeuwen,
  van~der Marel, Mignard, Murray, Le~Poole, Schrijver, Turon, Arenou,
  Froeschl{\'e}, \& Petersen}]{1997AA...323L..49P}
Perryman, M. A.~C., Lindegren, L., Kovalevsky, J., {et~al.} 1997, Astronomy and
  Astrophysics 323, 323, L49

\bibitem[{Pesch(1968)}]{1968ApJ...151..605P}
Pesch, P. 1968, The Astrophysical Journal, 151, 605

\bibitem[{Phan~Bao {et~al.}(2008)Phan~Bao, Bessell, Mart{\'\i}n, Simon,
  Borsenberger, Tata, Guibert, Crifo, Forveille, Delfosse, Lim, \&
  de~Batz}]{2008MNRAS.383..831P}
Phan~Bao, N., Bessell, M.~S., Mart{\'\i}n, E.~L., {et~al.} 2008, Monthly
  Notices of the Royal Astronomical Society, 383, 831

\bibitem[{Platais {et~al.}(1998)Platais, Kozhurina-Platais, \& van
  Leeuwen}]{1998AJ....116.2423P}
Platais, I., Kozhurina-Platais, V., \& van Leeuwen, F. 1998, The Astronomical
  Journal, 116, 2423

\bibitem[{P{\"o}hnl \& Paunzen(2010)}]{2010AA...514A..81P}
P{\"o}hnl, H., \& Paunzen, E. 2010, Astronomy {\&} Astrophysics, 514, A81

\bibitem[{Rameau {et~al.}(2013)Rameau, Chauvin, Lagrange, Boccaletti, Quanz,
  Bonnefoy, Girard, Delorme, Desidera, Klahr, Mordasini, Dumas, \&
  Bonavita}]{2013ApJ...772L..15R}
Rameau, J., Chauvin, G., Lagrange, A.-M., {et~al.} 2013, The Astrophysical
  Journal Letters, 772, L15

\bibitem[{Reid {et~al.}(2007)Reid, Cruz, \& Allen}]{2007AJ....133.2825R}
Reid, N.~I., Cruz, K.~K., \& Allen, P.~R. 2007, The Astronomical Journal, 133,
  2825

\bibitem[{Reid {et~al.}(1995)Reid, Gizis, \& Hawley}]{1995AJ....110.1838R}
Reid, N.~I., Gizis, J.~E., \& Hawley, S.~L. 1995, The Astronomical Journal,
  110, 1838

\bibitem[{Reid {et~al.}(2004)Reid, Cruz, Lowrance, Allen, Mungall, Kilkenny,
  Liebert, Hawley, Fraser, Covey, Kirkpatrick, \&
  Burgasser}]{2004AJ....128..463R}
Reid, N.~I., Cruz, K.~K., Lowrance, P., {et~al.} 2004, The Astronomical
  Journal, 128, 463

\bibitem[{Riaz {et~al.}(2006)Riaz, Gizis, \& Harvin}]{2006AJ....132..866R}
Riaz, B., Gizis, J.~E., \& Harvin, J. 2006, The Astronomical Journal, 132, 866

\bibitem[{Ricker {et~al.}(2015)Ricker, Winn, Vanderspek, Latham, Bakos, Bean,
  Berta-Thompson, Brown, Buchhave, Butler, Butler, Chaplin, Charbonneau,
  Christensen-Dalsgaard, Clampin, Deming, Doty, De~Lee, Dressing, Dunham, Endl,
  Fressin, Ge, Henning, Holman, Howard, Ida, Jenkins, Jernigan, Johnson,
  Kaltenegger, Kawai, Kjeldsen, Laughlin, Levine, Lin, Lissauer, MacQueen,
  Marcy, McCullough, Morton, Narita, Paegert, Pall{\'e}, Pepe, Pepper,
  Quirrenbach, Rinehart, Sasselov, Sato, Seager, Sozzetti, Stassun, Sullivan,
  Szentgyorgyi, Torres, Udry, \& Villasenor}]{2015JATIS...1a4003R}
Ricker, G.~R., Winn, J.~N., Vanderspek, R., {et~al.} 2015, Journal of
  Astronomical Telescopes, 1, 014003

\bibitem[{Riello {et~al.}(2018)Riello, De~Angeli, Evans, Busso, Hambly,
  Davidson, Burgess, Montegriffo, Osborne, Kewley, Carrasco, Fabricius, Jordi,
  Cacciari, van Leeuwen, \& Holland}]{Riello:2018bo}
Riello, M., De~Angeli, F., Evans, D.~W., {et~al.} 2018, Astronomy {\&}
  Astrophysics, doi:10.1051/0004-6361/201832712

\bibitem[{Rodriguez {et~al.}(2011)Rodriguez, Bessell, Zuckerman, \&
  Kastner}]{2011ApJ...727...62R}
Rodriguez, D., Bessell, M.~S., Zuckerman, B., \& Kastner, J.~H. 2011, The
  Astrophysical Journal, 727, 62

\bibitem[{Rodriguez {et~al.}(2013)Rodriguez, Zuckerman, Kastner, Bessell,
  Faherty, \& Murphy}]{2013ApJ...774..101R}
Rodriguez, D., Zuckerman, B., Kastner, J.~H., {et~al.} 2013, The Astrophysical
  Journal, 774, 101

\bibitem[{Royer {et~al.}(2007)Royer, Zorec, \& G{\'o}mez}]{2007AA...463..671R}
Royer, F., Zorec, J., \& G{\'o}mez, A.~E. 2007, Astronomy and Astrophysics,
  463, 671

\bibitem[{Sartoretti {et~al.}(2018)Sartoretti, Katz, Cropper, Panuzzo,
  Seabroke, Viala, Benson, Blomme, \& {et al.}}]{Sartoretti:2018jm}
Sartoretti, P., Katz, D., Cropper, M., {et~al.} 2018, Astronomy {\&}
  Astrophysics, doi:10.1051/0004-6361/201832836

\bibitem[{Saumon \& Marley(2008)}]{2008ApJ...689.1327S}
Saumon, D., \& Marley, M.~S. 2008, The Astrophysical Journal, 689, 1327

\bibitem[{Schlieder {et~al.}(2012)Schlieder, L{\'e}pine, Rice, Simon, Fielding,
  \& Tomasino}]{2012AJ....143..114S}
Schlieder, J.~E., L{\'e}pine, S., Rice, E., {et~al.} 2012, The Astronomical
  Journal, 143, 114

\bibitem[{Schmidt {et~al.}(2007)Schmidt, Cruz, Bongiorno, Liebert, \&
  Reid}]{2007AJ....133.2258S}
Schmidt, S.~J., Cruz, K.~K., Bongiorno, B.~J., Liebert, J., \& Reid, N.~I.
  2007, The Astronomical Journal, 133, 2258

\bibitem[{Schmidt {et~al.}(2015)Schmidt, Hawley, West, Bochanski, Davenport,
  Ge, \& Schneider}]{2015AJ....149..158S}
Schmidt, S.~J., Hawley, S.~L., West, A.~A., {et~al.} 2015, The Astronomical
  Journal, 149, 158

\bibitem[{Scholz {et~al.}(2005)Scholz, Meusinger, \&
  Jahrei{\ss}}]{2005AA...442..211S}
Scholz, R.~D., Meusinger, H., \& Jahrei{\ss}, H. 2005, Astronomy and
  Astrophysics, 442, 211

\bibitem[{Shkolnik {et~al.}(2017)Shkolnik, Allers, Kraus, Liu, \&
  Flagg}]{2017AJ....154...69S}
Shkolnik, E.~L., Allers, K.~N., Kraus, A.~L., Liu, M.~C., \& Flagg, L. 2017,
  The Astronomical Journal, 154, 69

\bibitem[{Shkolnik {et~al.}(2012)Shkolnik, Anglada-Escude, Liu, Bowler,
  Weinberger, Boss, Reid, \& Tamura}]{2012ApJ...758...56S}
Shkolnik, E.~L., Anglada-Escude, G., Liu, M.~C., {et~al.} 2012, The
  Astrophysical Journal, 758, 56

\bibitem[{Shkolnik {et~al.}(2011)Shkolnik, Liu, Reid, Dupuy, \&
  Weinberger}]{2011ApJ...727....6S}
Shkolnik, E.~L., Liu, M.~C., Reid, I.~N., Dupuy, T., \& Weinberger, A.~J. 2011,
  The Astrophysical Journal, 727, 6

\bibitem[{Silaj \& Landstreet(2014)}]{2014AA...566A.132S}
Silaj, J., \& Landstreet, J.~D. 2014, Astronomy and Astrophysics, 566, A132

\bibitem[{Skrutskie {et~al.}(2006)Skrutskie, Cutri, Stiening, Weinberg,
  Schneider, Carpenter, Beichman, Capps, Chester, Elias, Huchra, Liebert,
  Lonsdale, Monet, Price, Seitzer, Jarrett, Kirkpatrick, Gizis, Howard, Evans,
  Fowler, Fullmer, Hurt, Light, Kopan, Marsh, McCallon, Tam, Van~Dyk, \&
  Wheelock}]{2006AJ....131.1163S}
Skrutskie, M.~F., Cutri, R.~M., Stiening, R., {et~al.} 2006, The Astronomical
  Journal, 131, 1163

\bibitem[{Smart {et~al.}(2017)Smart, Marocco, Caballero, Jones, Barrado,
  Beam{\'\i}n, Pinfield, \& Sarro}]{2017MNRAS.469..401S}
Smart, R.~L., Marocco, F., Caballero, J.~A., {et~al.} 2017, Monthly Notices of
  the Royal Astronomical Society, 469, 401

\bibitem[{Soderblom {et~al.}(2014)Soderblom, Hillenbrand, Jeffries, Mamajek, \&
  Naylor}]{2014prpl.conf..219S}
Soderblom, D.~R., Hillenbrand, L.~A., Jeffries, R.~D., Mamajek, E.~E., \&
  Naylor, T. 2014, Protostars and Planets VI, 219

\bibitem[{Song {et~al.}(2012)Song, Zuckerman, \& Bessell}]{2012AJ....144....8S}
Song, I., Zuckerman, B., \& Bessell, M.~S. 2012, The Astronomical Journal, 144,
  8

\bibitem[{Soubiran {et~al.}(2018)Soubiran, Jasniewicz, Chemin, Zurbach,
  Brouillet, Panuzzo, Sartoretti, Katz, \& {et al.}}]{Soubiran:2018fz}
Soubiran, C., Jasniewicz, G., Chemin, L., {et~al.} 2018, Astronomy {\&}
  Astrophysics, doi:10.1051/0004-6361/201832795

\bibitem[{Stephenson(1986)}]{1986AJ.....92..139S}
Stephenson, C.~B. 1986, Astronomical Journal, 92, 139

\bibitem[{Torres {et~al.}(2006)Torres, Quast, da~Silva, de~La~Reza, Melo, \&
  Sterzik}]{2006AA...460..695T}
Torres, C. A.~O., Quast, G.~R., da~Silva, L., {et~al.} 2006, Astronomy and
  Astrophysics, 460, 695

\bibitem[{Torres {et~al.}(2008)Torres, Quast, Melo, \&
  Sterzik}]{2008hsf2.book..757T}
Torres, C. A.~O., Quast, G.~R., Melo, C. H.~F., \& Sterzik, M.~F. 2008,
  Handbook of Star Forming Regions, I, 757

\bibitem[{Upgren(1962)}]{1962AJ.....67...37U}
Upgren, A.~R. 1962, Astronomical Journal, 67, 37

\bibitem[{van Altena(1966)}]{1966AJ.....71..482V}
van Altena, W.~F. 1966, Astronomical Journal, 71, 482

\bibitem[{van Belle \& von Braun(2009)}]{2009ApJ...694.1085V}
van Belle, G.~T., \& von Braun, K. 2009, The Astrophysical Journal, 694, 1085

\bibitem[{West {et~al.}(2008)West, Hawley, Bochanski, Covey, Reid, Dhital,
  Hilton, \& Masuda}]{2008AJ....135..785W}
West, A.~A., Hawley, S.~L., Bochanski, J.~J., {et~al.} 2008, The Astronomical
  Journal, 135, 785

\bibitem[{West {et~al.}(2011)West, Morgan, Bochanski, Andersen, Bell, Kowalski,
  Davenport, Hawley, Schmidt, Bernat, Hilton, Muirhead, Covey, Rojas-Ayala,
  Schlawin, Gooding, Schluns, Dhital, Pineda, \& Jones}]{2011AJ....141...97W}
West, A.~A., Morgan, D.~P., Bochanski, J.~J., {et~al.} 2011, The Astronomical
  Journal, 141, 97

\bibitem[{Wilson(1953)}]{1953GCRV..C......0W}
Wilson, R.~E. 1953, Washington, 0

\bibitem[{Wright {et~al.}(2003)Wright, Egan, Kraemer, \&
  Price}]{2003AJ....125..359W}
Wright, C.~O., Egan, M.~P., Kraemer, K.~E., \& Price, S.~D. 2003, The
  Astronomical Journal, 125, 359

\bibitem[{Wright {et~al.}(2010)Wright, Eisenhardt, Mainzer, Ressler, Cutri,
  Jarrett, Kirkpatrick, Padgett, McMillan, Skrutskie, Stanford, Cohen, Walker,
  Mather, Leisawitz, Gautier, McLean, Benford, Lonsdale, Blain, Mendez, Irace,
  Duval, Liu, Royer, Heinrichsen, Howard, Shannon, Kendall, Walsh, Larsen,
  Cardon, Schick, Schwalm, Abid, Fabinsky, Naes, \& Tsai}]{2010AJ....140.1868W}
Wright, E.~L., Eisenhardt, P. R.~M., Mainzer, A.~K., {et~al.} 2010, The
  Astronomical Journal, 140, 1868

\bibitem[{Zuckerman {et~al.}(2006)Zuckerman, Bessell, Song, \&
  Kim}]{2006ApJ...649L.115Z}
Zuckerman, B., Bessell, M.~S., Song, I., \& Kim, S. 2006, The Astrophysical
  Journal, 649, L115

\bibitem[{Zuckerman \& Song(2004)}]{2004ARAA..42..685Z}
Zuckerman, B., \& Song, I. 2004, Annual Review of Astronomy {\&}Astrophysics,
  42, 685

\end{thebibliography}

\end{document}